\documentclass[12pt,a4paper]{iopart}

\usepackage{iopams}
\usepackage{epsfig}
\usepackage{graphicx}
\usepackage{amssymb}
\usepackage{amsfonts}
\usepackage{color}
\usepackage{psfrag}
\usepackage{sidecap}
\usepackage{fancyhdr}
\usepackage[breaklinks=true,colorlinks=true,linkcolor=blue,citecolor=red,filecolor=blue,urlcolor=blue]{hyperref}

\makeatletter
\newcommand\appendix@section[1]{%
  \refstepcounter{section}%
  \orig@section*{Appendix \@Alph\c@section: #1}%
  \addcontentsline{toc}{section}{Appendix: #1}%
}
\let\orig@section\section
\g@addto@macro\appendix{\let\section\appendix@section}
\makeatother

\sidecaptionvpos{figure}{c}

\begin{document}

\title{Tunneling and propagation of vacuum bubbles on dynamical backgrounds}

\author{Dennis Simon$\,^1$, Julian Adamek$\,^1$, Aleksandar Raki\'c$\,^1$ and \\ Jens C. Niemeyer$\,^{1,2}$}

\address{$^1$ Institut f\"ur Theoretische Physik und Astrophysik, Universit\"at W\"urzburg,\\ Am Hubland, D-97074 W\"urzburg, Germany\\
$^2$ Institut f\"ur Astrophysik, Universit\"at G\"ottingen, Friedrich-Hund-Platz 1,\\ D-37077 G\"ottingen, Germany}

\ead{dsimon@astro.uni-wuerzburg.de\\
jadamek@physik.uni-wuerzburg.de\\
rakic@astro.uni-wuerzburg.de\\
niemeyer@astro.physik.uni-goettingen.de}


\vspace{0.8cm}

\begin{abstract}
In the context of bubble universes produced by a first-order 
phase transition with large nucleation rates compared to the 
inverse dynamical time scale of the parent bubble, we extend 
the usual analysis to non-vacuum backgrounds. In particular, 
we provide semi-analytic and numerical results for the modified 
nucleation rate in FLRW backgrounds, as well as a parameter 
study of bubble walls propagating into inhomogeneous (LTB) or 
FLRW spacetimes, both in the thin-wall approximation. We show 
that in our model, matter in the  background often prevents 
bubbles from successful expansion and forces them to collapse. 
For cases where they do expand, we give arguments why the 
effects on the interior spacetime are small for a wide range
of reasonable parameters and discuss the limitations of the 
employed approximations.
\end{abstract}

\vspace{0.8cm}


\newpage
\tableofcontents

\section{Introduction}

In the recent literature a lot of interest has been dedicated to the 
question of how inflationary models \cite{Guth00,Linde07} can be embedded 
into a general theory \cite{McAllister07}. In this respect the emergence 
of the string theory landscape \cite{Susskind03} has opened various 
possibilities and a completely new perspective. Due
to flux compactification in string theory there can be a huge number of
distinct vacua \cite{Douglas06} in scalar field space without a unique
physical selection mechanism available at present.

Classically, a field that is stuck in one
vacuum would be trapped forever and
could never move through the landscape. Quantum mechanically, the vacua
are rendered metastable by the possibility for any field to tunnel to
another (metastable) vacuum and thereby
probe the landscape. The rate of tunneling naturally depends on
the energy difference between the vacua and on the height and width of
the potential barrier. The transition by pure tunneling can be 
described by the Coleman-De Luccia instanton \cite{CDL80} whereas a 
transition over the barrier due to thermal fluctuations can be described 
by the Hawking-Moss instanton \cite{HawkingMoss}.

\begin{figure}[t]
\psfrag{vphi}[t][t]{$V(\phi)$\hspace{-30pt}}
\psfrag{phi1}[t][t]{$\phi_1$\hspace{-16pt}}
\psfrag{phi2}[t][t]{$\phi_2$\hspace{-20pt}}
\psfrag{dS}[t][t]{dS}
\centerline{\includegraphics[width=0.84\textwidth]{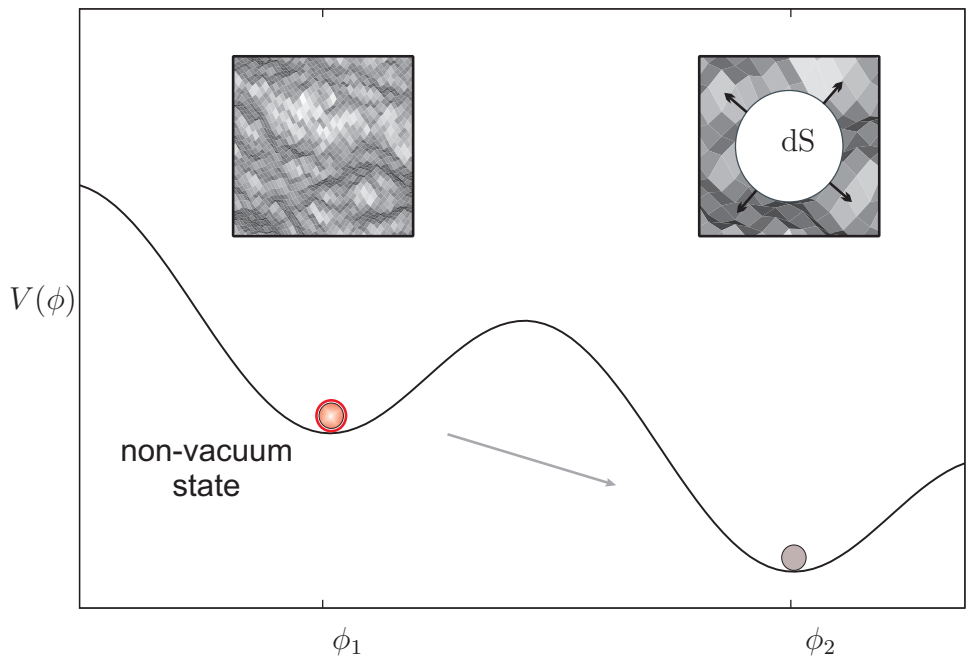}}
\caption{We investigate the effects on the tunneling and the 
evolution of bubbles of new vacuum when the precursor state is 
non-vacuum. The simplistic sketch shows such a tunneling event that 
can be imagined in the context of rapid tunneling in the landscape. 
We are interested in the case when tunneling is rapid enough such 
that the background had no time to relax to vacuum and is, for 
instance, undergoing a cosmological phase transition. This 
complication can be relevant in the context of chain inflation 
\cite{Freese05}--\cite{Ashoorioon08b} or other models 
that invoke rapid tunneling, like e.g.~DBI or resonance tunneling 
\cite{Sarangi07}, \cite{Tye06}. For some particular examples of non-vacuum 
backgrounds (like e.g.~radiation dominated FLRW or inhomogeneous 
LTB), we analyse the effects on the semiclassical tunneling rate 
as well as on the subsequent general relativistic evolution of 
the formed new vacuum bubbles.}
\label{fig:potential}
\end{figure}

The Coleman-De Luccia process gives rise to the nucleation of 
spherically symmetric regions (bubbles) in space which are filled 
with new vacuum and expand into the old vacuum 
-- a first-order phase transition. So, in the landscape multiple 
tunneling from multiple metastable vacua can occur, leading to 
different patches of spacetime undergoing coeval inflation in a 
variety of vacua. This process leads to a very complicated (fractal) 
large-scale structure of the universe that has been termed the 
attractor of eternal inflation. It is an attractor in the sense that 
statistical properties of the large-scale universe asymptotically do 
not depend on initial conditions \cite{Garriga05}.

The probability per unit four-volume for a tunneling to occur may be 
written as
\begin{equation}
\Gamma = A \exp\left(-2 \mathrm{Im}\mathcal{S}\right)~,
\end{equation}
where $A$ is a determinantal factor and $\mathcal{S}$ is the action of 
the instanton mediating the tunneling process. The prefactor $A$ is 
notoriously hard to obtain and is usually assumed to be of order 
unity for most practical purposes. The instanton action $\mathcal{S}$ 
can be obtained by solving the field equations with appropriate 
boundary conditions. In many cases the layer separating the two phases 
can be thought of as a domain wall with a thickness that is small 
compared to the size of the bubble (\lq thin-wall approximation\rq). 
With this assumption a calculation in Minkowski spacetime \cite{Coleman77} 
yields $\mathrm{Im}\mathcal{S} = 27 \pi^2 \sigma^4 / 4 \epsilon^3\,$, with 
$\epsilon$ the latent heat and $\sigma$ the energy density in the surface layer.
It is often assumed that $\sigma \gg \epsilon$ (in Planck units)
and thus tunneling is suppressed by a huge 
exponential factor, which implies extremely long lifetimes of the 
metastable vacua. However, any given point will eventually enter the 
decay chain, no matter what the odds are.

In this work we will study bubbles which nucleate from a non-vacuum 
initial state, that is inside pockets which do not obey the de~Sitter 
symmetries because they are not vacuum dominated at the time when 
tunneling takes place, see Fig.~\ref{fig:potential}. We look at 
tunneling from a precursor state that is, say, in a radiation 
or a matter dominated phase (but not yet vacuum dominated), and which 
produces an inflationary bubble of de~Sitter vacuum. One may object 
that a large class \cite{Wald83,Jensen86,Kitada92,Nobre09} of 
inflating spacetimes, if they include a positive cosmological 
constant, will always be vacuum dominated in the asymptotic future 
(the cosmic no-hair conjecture~\cite{HawkingMoss,Gibbons77,Starobinski83}).
Together with the aforementioned long lifetimes of metastable vacua 
in the landscape this seems to render the question of bubble 
nucleation from anything other than the vacuum academic. However, 
there are recent proposals of inflation where tunneling between 
minima on the landscape occurs \emph{rapidly}. This is the case, 
e.g., in the model of chain inflation\footnote{
In chain inflation, tunneling occurs stepwise through a sequence of 
many minima. Chain inflation 
resurrects the old inflationary idea of a first-order phase 
transition but is able to solve the problem of graceful exit. 
Originally, this was accomplished because the fields were assumed 
to be coupled. The coupling is responsible for rapid tunneling: 
once a first tunneling has occurred the field increases the decay 
probability of its neighbor(s) further up in the landscape due to 
coupling. A chain reaction of rapid tunneling starts that 
eventually ends with a (slow) transition to true vacuum.
However, chain inflation does not seem to allow for eternal 
inflation because it would not produce the right 
primordial density fluctuations \cite{Feldstein06}. In \cite{Freese06} 
a concrete realization of chain inflation on the string landscape that 
is driven by four form fluxes has been proposed.} \cite{Freese05,Watson06,Feldstein06,Freese06,Huang07,Chialva08a,Chialva08b,Ashoorioon08a,Ashoorioon08b}
or for non-standard tunneling on the landscape via DBI or 
resonance tunneling \cite{Sarangi07}, \cite{Tye06}. In case of 
rapid tunneling -- in chain inflation every transition only 
yields a fraction of an e-fold -- the efficiency of the cosmic 
no-hair mechanism can be questioned. Therefore, we argue that 
the question of the consequences of rapid inflation through a 
pocket that has possibly not yet relaxed to pure vacuum deserves 
some attention. We will attack this problem in a twofold way.
On the one hand, we will look at the tunneling itself and on the
other hand, we will study the classical evolution of a vacuum bubble
in a non-vacuum environment.

Concerning the former, one would for instance expect that the instanton 
action picks up some geometrical corrections due to the non-trivial 
background. Any such modifications are potentially interesting, 
since the tunneling rates are exponentially sensitive to them. We 
will show how one can assess the relevance of those 
effects, based on a comparison of characteristic time scales. 
Geometrical corrections become important if some dynamical time 
scale of the background is comparable to or smaller than the 
characteristic time scale of the tunneling process, which is given 
by the bubble's proper nucleation radius. This may happen either when 
the nucleation radius is very large -- but then tunneling rates are 
so small that the background geometry will relax to a de~Sitter 
stage long before transition occurs -- or when the background 
dynamics is characterized by some very short time scales.

The subsequent classical evolution of bubbles in 
such an environment will also be modified with respect to the 
vacuum case. We will study the 
propagation of vacuum bubbles in presence of i) homogeneously 
distributed matter, ii) matter with an inhomogeneous radial 
profile and iii) a fluid undergoing a second-order phase transition.
These simple toy scenarios shall give a taste of the phenomenology 
which may result from non-vacuum initial states.

Besides for chain inflation, the results of this analysis can be 
important also for other scenarios. For instance, the influence of 
background inhomogeneity can be relevant in the context of 
landscape sampling by tunneling. While through resonant 
processes the tunneling rate on the landscape can be enhanced 
\cite{Tye06}, inhomogeneous initial states may be of high 
importance for this sort of tunneling \cite{Saffin08,Copeland07}. 
The approach in our work is partly inspired by \cite{Fischler07}, 
where it was studied how an inhomogeneous and spherically 
sym\-metric background influences the classical evolution of an 
inflationary bubble of new vacuum immersed into it. In the limits of 
the used model it was claimed in \cite{Fischler07} that ambient 
inhomogeneities do not alter the evolution of bubbles 
significantly as long as the weak energy condition is 
respected. On the other hand, this topic also touches the 
interesting issue of inhomogeneous initial conditions versus the 
onset of inflation, see 
e.g.~\cite{Goldwirth91,Calzetta:1992gv,KurkiSuonio93,Deruelle94,Iguchi96,Berera00}.

The paper is organized as follows. In section 2 we present 
results that show the influence of dynamical backgrounds on the 
nucleation rate of bubbles of new vacuum. To this end, we apply an 
extension of the usual semiclassical approach to cosmologically 
interesting time-dependent FLRW backgrounds such as power law inflation 
or radiation dominated universes. The nucleation rates are obtained
in the thin-wall approximation by using a complex time path formalism.
In section 3, we analyze the subsequent classical trajectory of bubbles.
In addition to FLRW backgrounds\footnote{We thank Ben Freivogel for this suggestion.},
we look at the evolution in an exact 
spherically symmetric and inhomogeneous spacetime containing also matter.
The interior of the bubble is assumed to be de~Sitter spacetime,
as a first approximation to the inflationary phase 
of the patch of the universe that we are observing. The bubble evolution
follows from the Israel junction method which is employed to
join these spacetimes together. We ask whether signatures of the background
can potentially be seen by an observer inside the bubble.
In section 4 we summarize our results and give conclusions as well 
as some remarks on the limits of the used methods and an outlook. 
We use units $c = \hbar = G = 1$, and sign conventions for 
geometrical quantities in accordance with \cite{MTW}.

\section{Bubble nucleation in time-dependent settings}

Quantum nucleation of bubbles is a very intricate problem, especially when
effects of gravity have to be taken into account. Much of the literature
on this topic focusses on the case of transitions between two vacua having
different values of the cosmological constant. In this special case, a
semiclassical calculation of the nucleation rate based on instanton
methods has been presented by Coleman and De Luccia~\cite{CDL80}. However,
this calculation, as well as many alternative approaches developed by others (cf.~e.g.~\cite{FMP90}),
heavily relies on a high degree of symmetry of spacetime, which is
initially assumed to be in a pure vacuum state with the geometry of
de Sitter spacetime. We therefore feel that the applicability of these results
to cases where spacetime is not (or only approximately) in a vacuum state
is in need of some clarification (see also the discussion in \cite{Widrow91}).

A calculation of nucleation rates in arbitrary non-vacuum states, including
all possible effects, is clearly beyond our capabilities. We will instead
only take a small step away from the assumption of de Sitter symmetry and
consider Friedmann universes in general, of which de Sitter spacetime is only a
special case. We therefore retain many useful simplifications, in particular
the assumption of homogeneity of space (but not of spacetime!), and the
so-called \lq thin-wall approximation\rq. The cosmological expansion
of the universe, however, follows a non-trivial dynamical law (the Friedmann
equation), and we are interested in the effect of the \textit{time-dependent}
Hubble rate on the nucleation rate of bubbles, which will in turn itself become
\textit{time-dependent}.

To keep matters as simple as possible, we will consider the nucleation of a
spherical bubble of new phase, and we will assume that its shell -- the layer
which separates the new phase from the old -- is of negligible thickness
compared to the size of the bubble. This amounts to the
\lq thin-wall approximation\rq. The energy budget of the bubble consists of
latent heat (the difference between the energy densities of the two phases)
and surface tension. Throughout this section we will neglect gravitational backreaction of the bubble
onto the spacetime geometry since we are primarily interested in the effects
of cosmological expansion of the background, which is taken into account, and
the treatment of the full gravitational problem would introduce too many
additional complications.

\subsection{Lagrangean formulation}

Associated to the background spacetime is the
Friedmann-Lema\^itre-Robertson-Walker (FLRW) line element
\begin{equation}
\rmd s^2 = a^2\!\left(\eta\right) \left[-\rmd \eta^2 + \rmd r^2 + r^2 \rmd \Omega^2\right]~,
\end{equation}
where we have assumed flat spatial sections. The conformal
time $\eta$ is related to cosmological (proper) time $t$ by $\rmd t = a \rmd \eta$.

Spherical symmetry reduces the bubble dynamics to a $1+1$ dimensional problem.
Denoting the coordinate radius of the shell as $\bar{r}$, the shell trajectory
$\bar{r}\left(\eta\right)$ follows from the action
\begin{equation}
\label{FRWaction}
\mathcal{S} = \int\!\rmd\eta \left[\frac{4 \pi}{3} \epsilon~a^4\!\left(\eta\right) \bar{r}^3\!\left(\eta\right) - 4 \pi \sigma~a^3\!\left(\eta\right) \bar{r}^2\!\left(\eta\right) \sqrt{1 - \left(\partial_\eta \bar{r}\left(\eta\right)\right)^2}\right]~,
\end{equation}
where $\epsilon$ and $\sigma$ denote, respectively, the difference between the
energy densities of the two phases (latent heat) and the surface energy density
(surface tension) of the shell. As we have indicated above, this effective action
does not take into account gravitational self-interaction of the bubble. In
order to include some of these effects, one could add surface-surface,
volume-volume, as well as surface-volume terms for gravitational energy.

The evolution of the scale factor $a$ introduces an explicit time-dependence,
giving rise to nucleation rates which will in general be time-dependent
as well. A formalism for calculating semiclassical tunneling rates in
time-dependent settings has been presented by Keski-Vakkuri and Kraus \cite{KVK96}.
Its application to the present scenario will be worked out in detail in the
following section.

\subsection{The complex time path formalism}

Our starting point is the classical equation of motion of the bubble, which
can be found from eq.~(\ref{FRWaction}) as
\begin{equation}
\label{r-eom}
4 \pi \epsilon a^4 \bar{r}^2 - 8 \pi \sigma a^3 \bar{r} \sqrt{1 - \left(\partial_\eta \bar{r}\right)^2} = \frac{\rmd}{\rmd \eta} \left[4 \pi \sigma \frac{a^3 \bar{r}^2 \partial_\eta \bar{r}}{\sqrt{1 - \left(\partial_\eta \bar{r}\right)^2}}\right]~,
\end{equation}
and we have dropped some labels in favor of simplified notation.

The classical trajectory after tunneling
emanates from a classical turning point, where the canonical momentum
\begin{equation}
\bar{p}~\equiv~\frac{\partial \mathcal{L}}{\partial \partial_\eta \bar{r}} = 4 \pi \sigma \frac{a^3 \bar{r}^2 \partial_\eta \bar{r}}{\sqrt{1 - \left(\partial_\eta \bar{r}\right)^2}}
\end{equation}
vanishes. It has been pointed out in \cite{KVK96} that by analytic
continuation to complex $\eta$ one can find a classical trajectory\footnote{It
turns out to be a special feature of time-dependent settings that this
trajectory is not along a purely imaginary direction as would be the
case in static settings, where, as a consequence, Euclideanization of
time is a valid prescription.} (in the complex $\eta$ plane) that smoothly
shrinks the bubble to zero size. To this end, it is useful to rewrite
eq.~(\ref{r-eom}) as an equation for $\eta\left(\bar{r}\right)$:
\begin{equation}
4 \pi \epsilon a^4 \bar{r}^2 \partial_{\bar{r}} \eta - 8 \pi \sigma a^3 \bar{r} \sqrt{\left(\partial_{\bar{r}} \eta\right)^2 - 1} = \frac{\rmd}{\rmd \bar{r}} \left[4 \pi \sigma \frac{a^3 \bar{r}^2}{\sqrt{\left(\partial_{\bar{r}} \eta\right)^2 - 1}}\right]~,
\end{equation}
which, after some simplification, becomes
\begin{equation}
\label{eta-ode}
\frac{\epsilon}{\sigma} a \sqrt{\left(\partial_{\bar{r}} \eta\right)^2 - 1} = 2 \frac{\partial_{\bar{r}} \eta}{\bar{r}} + 3 \frac{\partial_\eta a}{a} - \frac{\partial^2_{\bar{r}} \eta}{\left(\partial_{\bar{r}} \eta\right)^2 - 1}~.
\end{equation}

We are looking for the solution with the boundary conditions
\begin{equation}
\label{bc}
\bar{p}\left(\eta_0\right) = \left. 4 \pi \sigma \frac{a^3 \bar{r}^2}{\sqrt{\left(\partial_{\bar{r}} \eta\right)^2 - 1}}\right|_{\eta = \eta_0} = 0~,\qquad\qquad\partial_{\bar{r}}\eta\left(0\right) = 0~.
\end{equation}
The first condition matches the solution to the turning point, from which on it
remains on the real $\eta$ axis for increasing $\bar{r}$. The second condition
guarantees that the full spherically symmetric solution is regular at the origin
$\bar{r} = 0$. Since $\eta$ will depart from the real axis for radii
smaller than the nucleation radius,
it is clear that  one also has to analytically continue the
time-dependent scale factor $a$ to the complex plane. These boundary conditions do
not determine the solution entirely: we are still free to choose the \lq nucleation
time\rq~$\eta_0$. The \lq nucleation radius\rq, i.e.~the coordinate radius
of the bubble at the classical turning point, will accordingly be denoted
as $\bar{r}_0$, and fulfills $\eta\left(\bar{r}_0\right) = \eta_0$. This means that
we have a one-parameter family of solutions labeled by their individual nucleation
times. For each solution, the semiclassical tunneling rate is determined by
the imaginary part of its action:
\begin{equation}
\Gamma\left(\eta_0\right) \sim \exp\left[-2 \mathrm{Im} \mathcal{S}\left(\eta_0\right)\right]~,
\end{equation}
where it is again useful to write $\mathrm{Im} \mathcal{S}\left(\eta_0\right)$
in terms of $\eta\left(\eta_0; \bar{r}\right)$:
\begin{eqnarray}
\label{ImS}
\mathrm{Im} \mathcal{S}\left(\eta_0\right)&=&\mathrm{Im} \int_0^{\bar{r}_0}\!\!\rmd \bar{r} \left[\frac{4 \pi}{3} \epsilon a^4\!\left(\eta\left(\eta_0; \bar{r}\right)\right) \bar{r}^3 \partial_{\bar{r}} \eta\left(\eta_0; \bar{r}\right)\right.\nonumber\\
&&\hspace{75pt}\left.- 4 \pi \sigma a^3\!\left(\eta\left(\eta_0; \bar{r}\right)\right) \bar{r}^2 \sqrt{\left(\partial_{\bar{r}} \eta\left(\eta_0; \bar{r}\right)\right)^2 - 1} \right]
\end{eqnarray}

We have now all the tools to compute time-dependent tunneling rates.
Let us now turn to some explicit examples.

\subsection{Minkowski spacetime}

Before dealing with time-dependent backgrounds, let us briefly review the
situation in Minkowski spacetime. We can simply set $a = 1$ and $\eta = t$.
Since we know that the solution has to be invariant under boosts, a
natural ansatz for the trajectory is a hyperbola
\begin{equation}
\bar{r}^2 - \left(\eta - \eta_0\right)^2 = \bar{r}_0^2~,
\end{equation}
which yields
\begin{equation}
\label{Mink-sol}
\eta\left(\eta_0; \bar{r}\right) = \eta_0 + \sqrt{\bar{r}^2 - \bar{r}_0^2}~,
\end{equation}
where the positive sign is chosen for the square root, corresponding
to an expanding bubble for $\bar{r} > \bar{r}_0$. Inserting into eq.~(\ref{eta-ode})
one infers
\begin{equation}
\label{Mink-nucr}
\bar{r}_0 = \frac{3 \sigma}{\epsilon}~,
\end{equation}
and one can check that this solution fulfills the boundary conditions
(\ref{bc}). The imaginary part of its action can be found readily
from eq.~(\ref{ImS})~,
\begin{equation}
\label{ImS-Mink}
\mathrm{Im}\mathcal{S}_{\mathrm{Mink}} = \frac{\pi^2}{12} \epsilon \bar{r}_0^4 = \frac{27 \pi^2 \sigma^4}{4 \epsilon^3}~,
\end{equation}
which is exactly Coleman's result \cite{Coleman77}.

\subsection{de Sitter spacetime}

Our first example of an expanding universe will be de Sitter spacetime.
Although it can actually be written in static coordinates, the FLRW
metric with \textit{flat} spatial sections has a scale factor which
grows exponentially with time, $a = \exp\left(H t\right)$. Written
in conformal time this becomes $a = -1 / H \eta$, where $\eta$ runs
from $-\infty$ to $0$ as $t$ runs from $-\infty$ to $+\infty$. In
this case, eq.~(\ref{eta-ode}) reads
\begin{equation}
\label{eta-ode-dS}
-\frac{\epsilon}{\sigma H \eta} \sqrt{\left(\partial_{\bar{r}} \eta\right)^2 - 1} = 2 \frac{\partial_{\bar{r}} \eta}{\bar{r}} - \frac{3}{\eta} - \frac{\partial_{\bar{r}}^2 \eta}{\left(\partial_{\bar{r}} \eta\right)^2 - 1}~.
\end{equation}

While finding the complete solution to this problem does not seem
promising, it turns out that we can guess the relevant solution by
sensibly generalizing eq.~(\ref{Mink-sol}). Since de Sitter spacetime
(in flat coordinates) has a constant expansion rate $H$, we expect
that the \textit{proper} nucleation radius should be independent of
time. Now, if $\bar{r}_0$ is to be a \textit{comoving} radius, it
has to be divided by the scale factor, i.e.~instead of
eq.~(\ref{Mink-nucr}) we expect
\begin{equation} \label{dS_exactsolution1}
\bar{r}_0 = a^{-1}\!\left(\eta_0\right) \frac{3 \sigma}{\epsilon} = -H \eta_0 \frac{3 \sigma}{\epsilon}~.
\end{equation}

Remarkably, choosing this nucleation radius is enough to have eq.~(\ref{Mink-sol})
solve eq.~(\ref{eta-ode-dS}) with the boundary conditions (\ref{bc}).

The integral of eq.~(\ref{ImS}) is still solvable, and its imaginary
part is found to be
\begin{eqnarray}
\label{ImS-dS}
\mathrm{Im} \mathcal{S}_{\mathrm{dS}}&=&\frac{\pi^2 \epsilon}{3 H^4} \frac{\left(1 - \sqrt{1 + \left(3 H \sigma / \epsilon\right)^2}\right)^2}{\sqrt{1 + \left(3 H \sigma / \epsilon\right)^2}}\nonumber\\
&=&\frac{4 \pi^2 \epsilon}{3 H^4} \sinh^2\!\frac{1}{4} \ln\left(1 + \left(3 H \sigma / \epsilon\right)^2\right)~,
\end{eqnarray}
which is independent of the choice $\eta_0$ for the time of nucleation.
This means that the nucleation rate in de Sitter spacetime is time independent,
which is a manifestation of the fact that de Sitter spacetime has no true
dynamics. This expression reduces to the result of Minkowski spacetime 
eq.~(\ref{ImS-Mink}) in the limit of $H \rightarrow 0$. The first correction 
is of order $H^2$ and complies with the expansion obtained by Abbott, Harari 
and Park~\cite{AHP87}. We can also take the limit $\epsilon \rightarrow 0$, 
which corresponds to the nucleation of a domain wall separating two degenerate 
vacua. One finds
\begin{equation}
\lim_{\epsilon \rightarrow 0} \mathrm{Im} \mathcal{S}_{\mathrm{dS}} = \frac{\pi^2 \sigma}{H^3}~,
\end{equation}
in complete agreement with a result obtained by Basu, Guth and Vilenkin~\cite{BGV91}.

\subsection{More general FLRW spacetimes}

Thus far we have only considered static spacetimes in order to get some
experience using the new tools. Let us finally turn to more general
spacetimes, where the expansion rate is not assumed to be constant. A
simple deformation of de Sitter expansion is given by power law inflation,
where the scale factor grows as $a = \left(\eta_1 / \eta\right)^{1 + \alpha}$.
For small deformation parameters $\alpha$ this is an exact slow-roll
solution of inflation with a constant slow-roll parameter $-\partial_t H / H^2 \approx \alpha$.
$\eta_1$ denotes an (arbitrary) point in time where the scale factor is
normalized to unity. This spacetime is obviously not static, and we shall
use this simple example to study effects of time-dependent cosmological
expansion on tunneling rates. For power law inflation, eq.~(\ref{eta-ode})
reads
\begin{equation}
\frac{\epsilon}{\sigma} \left(\frac{\eta_1}{\eta}\right)^{1 + \alpha} \sqrt{\left(\partial_{\bar{r}} \eta\right)^2 - 1} = 2 \frac{\partial_{\bar{r}} \eta}{\bar{r}} - 3 \frac{1 + \alpha}{\eta} - \frac{\partial_{\bar{r}}^2 \eta}{\left(\partial_{\bar{r}} \eta\right)^2 - 1}~.
\end{equation}
We have been unable to find an analytic solution to this equation and
therefore decided to treat it numerically. Using the boundary conditions
(\ref{bc}) one can find numerical solutions for any choices of $\alpha$,
$\eta_0 / \eta_1$ and $\epsilon / \sigma$. A parameter study of the tunneling rates
reveals the following picture. There are now three different time scales
in the problem. The inverse expansion rate $H^{-1}\left(\eta\right)$ gives the time scale
on which the background (scale factor) changes significantly. But there is
now another time scale related to the change of the expansion rate itself
(higher order derivatives of the expansion rate are zero in power
law inflation). These two time scales have to be compared to the
\lq bubble crossing time\rq , which we define by the nucleation radius
divided by the speed of light, or roughly $3 \sigma / \epsilon$ in our units.

\begin{SCfigure}[5][tp]
\psfrag{ImS}[b][b]{\small $\mathrm{Im}\mathcal{S}~/~\mathrm{Im}\mathcal{S}_\mathrm{dS}$}
\psfrag{epsilon}[t][t]{\small $\left| \partial_t H / H^2 \right|$}
\psfrag{proper}[l][l]{\small proper time average}
\psfrag{conformal}[l][l]{\small conformal time average}
\centering
\includegraphics[width=0.55\textwidth]{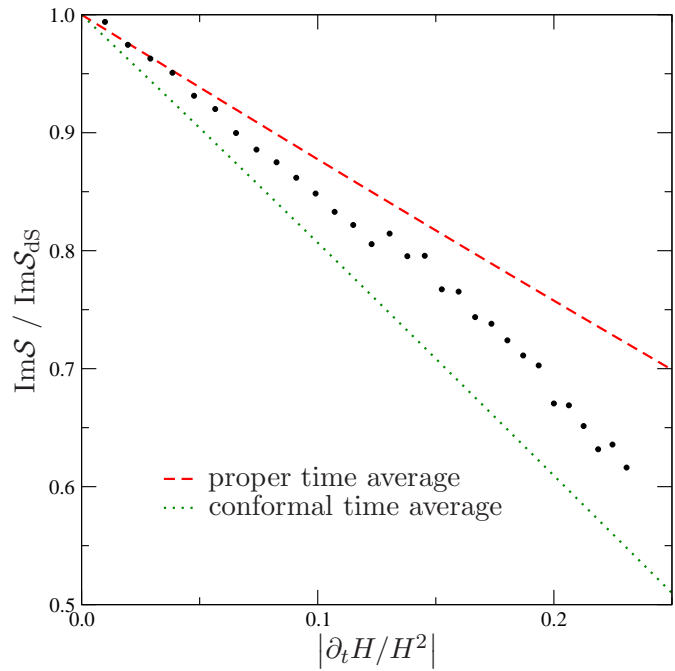}
\caption{\label{figpl} Numerical values for the imaginary part of the action $\mathcal{S}$ for
tunneling solutions with proper nucleation radius $3 \sigma / \epsilon = 2 H^{-1}\!\left(\eta_0\right)$,
as a function of the slow roll parameter $\partial_t H / H^2$. The values are normalized to the \lq quasistatic
approximation\rq, which is obtained from eq.~(\ref{ImS-dS}) by setting $H = H\left(\eta_0\right)$.
Thus, in this approximation only instantaneous dynamical parameters are taken into account. The tunneling
process, however, has a characteristic time scale given by the light-travel time across the bubble.
Hence it is more appropriate to take into account some \textit{average} dynamics of the background.
In a crude way, this can be accomplished by using an averaged expansion rate to evaluate the tunneling
probability in the quasistatic approximation. The
dashed red line is obtained from a \textit{proper time average} of $H$ taken over an interval
$\Delta t = 3 \sigma / \epsilon$ before nucleation, whereas the dotted green line is obtained from a
\textit{conformal time average} taken over an interval $\Delta \eta = \Delta t / a\left(\eta_0\right)$.}
\end{SCfigure}

If the bubble crossing time is the smallest time scale of the problem,
the tunneling rate is well approximated by the result of Minkowski spacetime,
eq.~(\ref{ImS-Mink}). However, if the bubble crossing time is not much smaller
than the Hubble time at nucleation, $H^{-1}\!\left(\eta_0\right)$, there are
two different possibilities. Either the characteristic time scale
for the \textit{change} of the expansion rate, $\left|\partial_t H / H\right|^{-1}$, is still much
larger than the bubble crossing time -- then we are in a regime where a quasistatic
approximation is valid such that a good estimate for the tunneling rate can be
obtained from eq.~(\ref{ImS-dS}) by setting $H = H\left(\eta_0\right)$. Or the
bubble crossing time cannot be regarded as small with respect to any other time scale. In
this case, the tunneling process \lq feels\rq~the changing expansion rate, and the
decay rate is modified significantly. Our numerical study clearly indicates that
the tunneling rate is enhanced compared to the quasistatic estimate. We believe
that this is related to the fact that the expansion rate decreases with time.
Instead of setting $H = H\left(\eta_0\right)$ in eq.~(\ref{ImS-dS}), one should
\textit{average} the expansion rate over an interval of one bubble crossing time
prior to nucleation. Using this averaged expansion rate in eq.~(\ref{ImS-dS})
turns out to give a much better estimate for the actual tunneling rate, cf.~Fig.~\ref{figpl}.

\subsection{Spacetimes with Big Bang singularity}

An interesting case which is also relevant for the chain inflation scenario is
a universe filled with radiation and some vacuum energy, such that radiation
dominates its early evolution, while it approaches a vacuum de Sitter phase at
later time. If we neglect the brief era of matter domination, this is also
a good model for our present universe (after reheating has taken place).
During the radiation era, the Hubble rate drops rapidly, and
assuming the vacuum is metastable, one might wonder if this affects tunneling
rates as described above. We can answer this question by comparing the relevant
time scales. If the universe is sufficiently flat, $a \propto t^{1/2}$
as long as radiation dominates over vacuum energy. Note that the universe has a Big
Bang singularity at $t = 0$ as long as reheating and any kind of cosmology before
the radiation era is not taken into account.
Hence, there exists a particle horizon of size $\sim t$.
Also both the Hubble time and the characteristic time scale for the change of the Hubble
rate are of order $\sim t$. Since even tunneling cannot violate causality, bubbles
larger than the particle horizon cannot be produced, at least not in the semiclassical
picture we are using here. This means that the bubble crossing time can never
be much larger than the other relevant time scales, and thus we expect
corrections to the tunneling rates to be small in general.
Moreover, a numerical study indicates that the tunneling rates for bubbles whose
nucleation radius is comparable to the size of the particle horizon become sensitive
to the dynamics of the scale factor up to the vicinity of the Big Bang, such that
details of reheating and even earlier cosmology become relevant.

As soon as vacuum energy begins to dominate, the time scale on which the expansion
rate changes tends to infinity quite rapidly as the Hubble rate approaches a constant.
Hence, for the late part of the evolution, we expect the de Sitter approximation
to be good. Note, however, that a particle horizon still exists and that bubbles should
not exceed the horizon size at any time as long as causal physics is at play. To
eliminate this \lq horizon problem\rq\footnote{Actually, this is nothing but the good
old horizon problem of standard Big Bang cosmology in a new guise, and therefore it can
be solved in the same spirit.}, one has to change the details of the model
near the Big Bang singularity to include any cosmology preceding the era of radiation
domination.

\section{Bubble propagation on dynamical backgrounds}

In this section we will explore the effect of background irregularities on the evolution of the bubble and whether these effects
can potentially be seen by an observer inside the bubble.
To this end we will follow a somewhat different route than in the preceding section,
where we ignored gravitational backreaction of the bubble.
In order to include this backreaction, we will follow the approach used in \cite{Fischler07} and assume that the bubble wall separates spacetime into two parts,
described by different metrics and containing different matter.
These two parts, for convenience called interior and exterior part,
are approximated by the manifolds $\mathcal{M}_-$ and $\mathcal{M}_+$ and are joined along a common timelike,
spherically symmetric hypersurface $\Sigma$ which represents the bubble wall.
Since the work of Israel \cite{Israel} there exists a well established formalism for joining manifolds along a common boundary,
known as the Israel junction conditions.
Once $\mathcal{M}_-$ and $\mathcal{M}_+$ are given, the evolution of the bubble wall uniquely follows from these conditions. Moreover, by this construction, the resulting spacetime is a solution to Einstein's field equations.

For the interior part we use de~Sitter spacetime (which can be thought of as an approximation to the inflationary phase of our observable patch of the universe) and for the exterior part two different cases will be considered.
We will study the evolution of vacuum bubbles in an inhomogeneous background for which the exterior part is approximated by the spherically symmetric Lema\^ itre-Tolman-Bondi (LTB) spacetime \cite{Lemaitre,Tolman,Bondi}.
To maintain spherical symmetry it is assumed that the bubble nucleates in the center of the LTB model.
In a similar vein, we will also explore the evolution of vacuum bubbles when the exterior part is given by a FLRW universe filled with a fluid which at some time undergoes a smooth (second-order) phase transition, e.g.~from $w=-1$ to $w=1/3$, or vice versa.

In the first subsection we will introduce the description of the bubble wall and the interior and exterior parts of spacetime.
Thereafter we will give a concise guide on how to calculate the junction equations and write them down explicitly for the cases of our interest. Finally we will solve these equations numerically and discuss the results.

\subsection{Bubble wall and background spacetime}

\subsubsection{Interior: de~Sitter}

The interior of the bubble is assumed to be in a de~Sitter phase with vacuum energy density given by $\Lambda/(8\pi)$.
We employ the flat slicing in which the metric is given by
\begin{equation}
  {\rmd s}^2 = -{\rmd t}^2 +\exp\left( 2\sqrt{\Lambda/3}~t\right)\left({\rmd r}^2 +r^2{\rmd\Omega}^2\right)~,
\end{equation}
and stress-energy is given by $T_{\mu\nu} = -\Lambda g_{\mu\nu}$.
All quantities should carry the index $(-)$ which indicates that they belong to $\mathcal{M}_-$.
For convenience we write these indices only when necessary,
hoping that it will be clear from the context which quantities are meant.

\subsubsection{Bubble wall}

The timelike, spherically symmetric hypersurface separating the interior and exterior parts of spacetime, the bubble wall,
is described by the metric
\begin{equation}\label{metric_bubble}
  {\rmd s}^2 = -{\rmd\tau}^2 + R^2{\rmd\Omega}^2~.
\end{equation}
Though it has been shown~\cite{BGG} that the stress-energy of a wall which separates regions of different vacua will solely be given by the surface tension, it is not clear whether this conclusion is valid when matter is present.
However, in lack of a field theoretic description, we assume that stress-energy is given by
\begin{equation}\label{seTshell}
  S_{ij} = -\sigma h_{ij}~.
\end{equation}
where $h_{ij}$ is the metric tensor of~(\ref{metric_bubble}).
The equations of motion of the proper radius $R$ and surface tension $\sigma$ of the bubble will be given by the junction conditions.

\subsubsection{Exterior: LTB and FLRW}

\paragraph{LTB spacetime}

In order to explore the evolution of a bubble in an inhomogeneous background we use an LTB ansatz.
In comoving coordinates a suitable metric \cite{PK} can be given in the form
\begin{equation} \label{metricLTB}
  {\rmd s}^2 = -{\rmd t}^2 + \frac{\left(r\partial_r a(t,r) + a(t,r)\right)^2}{1+2E(r)}{\rmd r}^2 + a^2(t,r)r^2{\rmd\Omega}^2~,
\end{equation}
with $2E(r)>-1$ but otherwise arbitrary.
From Einstein's equations with a dust source
$T_{\mu\nu} = \rho\delta_\mu^t\delta_\nu^t -\Lambda g_{\mu\nu}\,$, we obtain the equation of motion of the scale factor
\begin{equation} \label{eqmoLTB}
  \left(\frac{\partial_t a}{a}\right)^2 - \frac{2E}{a^2r^2} = \frac{2M}{a^3r^3} +\frac{\Lambda}{3}~.
\end{equation}
Here, $M(r)$ is the first integral of motion which corresponds to the active gravitational mass within a sphere of coordinate radius $r$.
Once the scale factor is known, the dust density $\rho$ is determined by
\begin{equation} \label{eqmodust}
  8\pi\rho = \frac{2\partial_r M}{a^2r^2\left(r\partial_ra+a\right)}~.
\end{equation}

\paragraph{FLRW spacetime}
In addition, we want to study the motion of bubbles in a background which undergoes a phase transition.
Therefore a flat FLRW spacetime is employed, a comoving coordinate system of which is
\begin{equation}
  {\rmd s}^2 = -{\rmd t}^2 +a^2(t)\left({\rmd r}^2 +r^2{\rmd\Omega}^2\right)~,
\end{equation}
with stress-energy given by
\begin{equation}
  T_{\mu\nu} = \left(\rho+p\right)\delta_\mu^t\delta_\nu^t + p g_{\mu\nu} -\Lambda g_{\mu\nu}~.
\end{equation}
The evolution of this background follows from the Friedmann equation
\begin{equation} \label{eqmoFLRW}
	\left(\frac{\partial_t a}{a}\right)^2 = \frac{8\pi}{3}\rho + \frac{\Lambda}{3}~,
\end{equation}
and continuity equation
\begin{equation} \label{conFLRW}
  \partial_t \rho +3 \frac{\partial_t a}{a}\left(\rho+p\right) = 0~.
\end{equation}
To look at the influence of a phase transition in the background on the evolution of the bubble,
we artificially\footnote{A universe with two or more components with different equations of state,
like e.g.~radiation and cosmological constant, actually has one or several intrinsic phase transitions.
However, these transitions are very gentle and would probably only produce a minuscule effect.}
introduce an abrupt change in the equation of state $p=w\rho$ via
\begin{equation}
  w(t) = -\frac{1}{3}\left(1 \pm 2\tanh(\gamma_\mathrm{pt} (t-t_\mathrm{pt})) \right)~.
\end{equation}
to model a nearly instantaneous (on time scale $\gamma_\mathrm{pt}^{-1} \ll H^{-1}$) phase transition at $t=t_\mathrm{pt}$ from $w=-1$ to $w=1/3$ (\lq reheating\rq)
and vice versa. Solutions to equations~(\ref{eqmoFLRW}) and~(\ref{conFLRW}) can not be given in closed form
and will be obtained numerically.

\subsection{Conditions for a valid junction}

The problem of joining manifolds across a common boundary has been lucidly explained in the work of Israel~\cite{Israel}.
For a nice introduction see also the textbooks~\cite{Poisson,GH}.
Two conditions arise in the course of joining two manifolds.
The first junction condition requires the induced metrics $h_{ij} =g_{\mu\nu} \mathrm{e}_i^\mu\mathrm{e}_j^\nu$ to coincide on $\Sigma$
\begin{equation} \label{junction1a}
	\left[h_{ij}\right] \equiv h_{ij}^+\vert_\Sigma -h_{ij}^-\vert_\Sigma = 0~.
\end{equation}
The second junction states that, whenever there is a discontinuity in the extrinsic curvature of $\Sigma$ as seen from
$\mathcal{M}_\pm$, a surface layer of stress-energy $S_{ij}$, given by
\begin{equation} \label{junction2a}
	8\pi S_{ij} = \left[K_{ij}\right] - h_{ij}\left[K\right]
\end{equation}
will be present.
Therefore the proposed stress-energy on the bubble wall~(\ref{seTshell})
has to be identified with the difference in the extrinsic curvature.
The components of the extrinsic curvature tensor $K_{ij}$ are defined as the covariant derivative of the vector
$\mathrm{e}_j^\mu$ along $\mathrm{e}_i^\nu$ projected onto the surface normal
\begin{equation}\label{ecT}
  K_{ij} =  n_\alpha\Gamma^\alpha_{\mu\nu}\mathrm{e}^\mu_i\mathrm{e}^\nu_j~.
\end{equation}
Once the projectors $\mathrm{e}_i^\mu = \partial x^\mu/\partial y^i$ are known,
the normal vector of $\Sigma$ can be obtained by the conditions
\begin{equation}
  n_\mu n^\mu = 1 \quad \mathrm{and}\quad n_\mu\mathrm{e}_i^\mu = 0
\end{equation}
up to a sign which determines how $\mathcal{M}_-$ and $\mathcal{M}_+$ are stuck together.
We choose this sign such that
\begin{equation}\label{sign_normal}
  n_\mu = \sqrt{g_{rr}}\left(-\dot r, \dot t, 0 ,0 \right)~,
\end{equation}
where a dot refers to a partial derivative with respect to $\tau$. This choice implies that in $\mathcal{M}_-$,
radii increase towards $\Sigma$ and in $\mathcal{M}_+$, radii decrease towards $\Sigma$.
The continuity equation
\begin{equation}\label{eqmosigma}
	\nabla_i S_j^i + \left[T^\alpha_\beta n_\alpha e^\beta_j\right] = 0~.
\end{equation}
is not independent of the two equations resulting from~(\ref{junction2a}) and will be used to substitute one of these.
We consider exterior stress-energy given by a perfect fluid and interior by a cosmological constant.
Therefore the $\tau$ component provides the following first order equation for $\sigma$
\begin{equation}\label{eqmosigma2}
  \dot\sigma = \left(\rho+p\right)\sqrt{g_{rr}}\dot r \dot t~.
\end{equation}
In the next subsections we will write down these equations for LTB spacetime.

\subsubsection{Equation of motion for the size and surface tension of the bubble}

By virtue of spherical symmetry of the bubble wall, angular coordinates can be identified
and the time and radial coordinate can be parameterized by the proper time of the bubble $\left(t(\tau),r(\tau)\right)$.
We write down the equations for the LTB part and relegate the FLRW equations to the appendix.
The conditions of the first junction turn out to be
\begin{equation} \label{junction1b}
	ar = R, \quad {\dot t}^2 = 1 + \frac{\left(r\partial_r a + a\right)^2}{1+2E}{\dot r}^2~.
\end{equation}
We will make use of the $\theta\theta$-component of the second junction condition which yields
\begin{equation} \label{junction2b}
  4\pi\sigma R = \sqrt{ {\dot R}^2 + 1-\frac{\Lambda_-}{3}R^2 } - \sqrt{ {\dot R}^2 + 1 -\frac{2M}{R} -\frac{\Lambda_+}{3}R^2 }~.
\end{equation}
Solving for the derivative we obtain the more convenient form
\begin{equation}\label{Roftau}
  \frac{1}{2}{\dot R}^2 + V = -\frac{1}{2}~,
\end{equation}
with $2V$ given by
\begin{equation}\hspace{-50pt}
	2V =
		-\left[ \frac{\Lambda_-}{3} +\left(\frac{\Lambda_+ -\Lambda_-}{24\pi\sigma} +2\pi\sigma\right)^2 \right]R^2
		-\left(1+\frac{\Lambda_+ -\Lambda_-}{48\pi^2\sigma^2}\right)\frac{M}{R}-\frac{M^2}{16\pi^2\sigma^2R^4}~.
\end{equation}
If $M=\mathrm{const}$ all coefficients in the potential are constant and we recover the Schwarzschild-de~Sitter model which has been discussed in~\cite{BGG,BKT,APS,AJ}.

However, since we want to introduce an exterior matter density, $M$ will no longer be constant, i.e. $M(r)=M(R/a)$ and the scale factor of the ambient spacetime will enter the equation.
In this way the motion of the surface becomes sensitive to the presence of matter in the background.

Note that~(\ref{metricLTB}) is covariant under a rescaling of the radial coordinate.
Together with $\partial_r M >0$ this allows one to define a radial coordinate such that
$M(r) = \frac{4\pi}{3}A r^3$ where $A$ is a constant. 
The potential becomes
\begin{equation}\label{VLTB}
  2V = -\left[\frac{\Lambda_-}{3}
       +\left(\frac{A}{3a^3\sigma} +\frac{\Lambda_+ -\Lambda_-}{24\pi\sigma} +2\pi\sigma \right)^2\right]R^2~.
\end{equation}
and the equation of motion for the surface tension is
\begin{equation}
  \dot\sigma = \rho\frac{r\partial_r a+a}{\sqrt{1+2E}}\dot r \dot t~.
\end{equation}
It is restricted by equation~(\ref{junction2b}) to
\begin{equation}\label{bound_sigma}
  4\pi\sigma < \sqrt{\frac{8\pi A}{3a^3} + \frac{\Lambda_+ -\Lambda_-}{3}} \equiv \sqrt{\frac{8\pi}{3}\epsilon}~.
\end{equation}
Here $\epsilon$ is the difference in energy density between inside and outside.
This bound is a direct consequence of the geometry that was fixed by the sign of the normal vector~(\ref{sign_normal}).

\subsubsection{Evolution equations expressed in exterior coordinates}

Since the background dynamics of the LTB (and FLRW) part of the spacetime can be obtained only numerically all equations will be solved in these coordinates in the first place.
Making use of the junction conditions we are able to write down the evolution equations in terms of the exterior coordinates.
After a solution to these equations has been obtained, the matching conditions will be employed again to express the evolution in terms of the interior coordinates. Like before, we explicitly write down the expressions for the LTB part and provide the FLRW equations in the appendix.

Let $\bar r(t)$ be the bubble radius in these coordinates.
Writing $\dot R = \dot t \frac{d}{dt}\left(a\bar r\right)$ we can solve for $\partial_t \bar r$ and obtain
\begin{equation}\label{posLTB}
  \partial_t \bar r =
    \frac{-(1+2E)\bar r\partial_t a +\sqrt{(1+2E)\left(1+2V\right)\left(\left(\bar r\partial_t a\right)^2-2E+2V\right)}}
    {\left(\bar r\partial_{\bar r} a+a\right)\left(2E-2V\right)}~.
\end{equation}
where we have chosen a positive sign of the square root because we are interested in solutions of physically growing bubbles,
i.e. $\dot R\geq 0$. The equation for $\sigma$~(\ref{eqmosigma2}) can as well be converted to LTB coordinates
\begin{equation} \label{sigmaLTB}
  \partial_t \sigma =
  \rho\frac{\left(\bar r\partial_{\bar r} a+a\right)\partial_t \bar r}
  {\sqrt{1+2E-\left(\bar r\partial_{\bar r} a+a\right)^2\left(\partial_t \bar r\right)^2}}~.
\end{equation}
The surface tension becomes time dependent.
It increases when the comoving radius of the bubble increases, i.e.~it collects matter from the background, and if the bubble shrinks it will exactly provide the amount of matter density determined by the background.
This is a limitation of the spacetime junction approach, which in the present form does not capture the
physics of matter transfer across the junction surface. Physically we would expect that dust would actually
penetrate into the bubble, as one can convince oneself by looking at the geodesics of \lq test particles\rq.
However, interior and exterior parts of spacetime are fixed ab initio and cannot be changed by the motion
of the bubble. Bearing with this limitation, we continue our analysis and will give an outlook
on this issue in our conclusions.

For now, equations~(\ref{posLTB}) and~(\ref{sigmaLTB}) completely determine the evolution of the bubble
in exterior coordinates.

\subsection{Bubble evolution on dynamical backgrounds}

In this section we explore the propagation of bubbles on dynamical backgrounds.
We start with an exact solution of a de~Sitter/de~Sitter spacetime and continue with the numerical solutions obtained for the
de~Sitter/LTB and de~Sitter/FLRW spacetimes.

\subsubsection{Exact solution in de~Sitter/de~Sitter spacetime}

\begin{SCfigure}[5][tp]
  \psfrag{Hr}[][]{\small $H\bar r$}
  \psfrag{Ht}[][]{\small $Ht$}
\centering
  \includegraphics[width=0.55\textwidth]{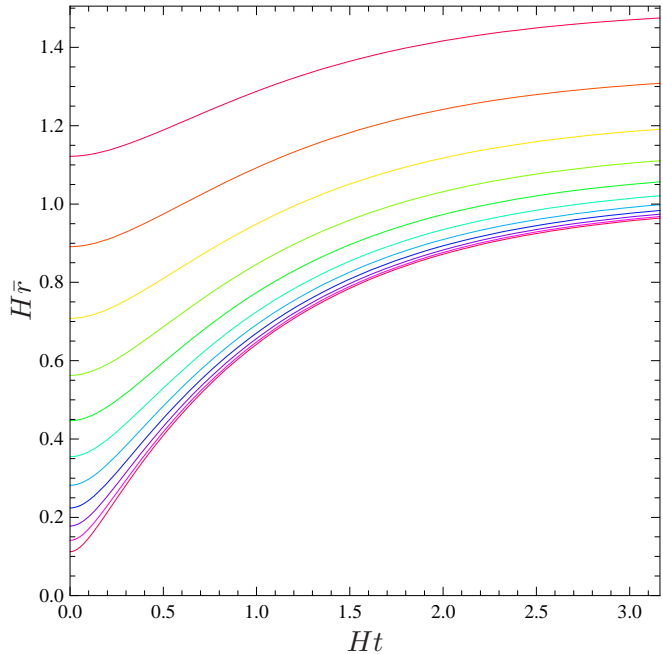}
  \caption{\label{dS_exact_trajectories2}
    Trajectories of the bubble wall in the flat slicing of de~Sitter/de~Sitter spacetime for different values of $u\,$; the
    parameter $u$ (\ref{def_of_u}) encodes the dependency on $\Lambda_\pm$ and the surface tension $\sigma$.
    The bubble expands accelerated from coordinate radius $\bar r_0=u^{-1}H^{-1}$ but converges to the finite coordinate radius
    $\sqrt{1+u^2}\bar r_0$ in the limit $t\rightarrow\infty$. The trajectory is shown in de~Sitter coordinates rather than
    in physical quantities (proper radius $R$ vs. proper time $\tau$) to make comparison to our later results easier.}
\end{SCfigure}

We solve equations~(\ref{posLTB}) and~(\ref{sigmaLTB}) for the case that both spacetimes are de~Sitter with cosmological constants given by $\Lambda_\pm$.
Then $M=0$ and we employ the flat slicing where also $E=0$. Since there is no matter in the background it follows from equation~ (\ref{sigmaLTB}) that $\sigma=\mathrm{const}$. The potential $V$ reduces to
\begin{equation}\label{VdS}
  2V = -\left[\frac{\Lambda_-}{3} +\left(\frac{\epsilon}{3\sigma} +2\pi\sigma\right)^2\right] R^2~.
\end{equation}
For further convenience we define
\begin{equation} \label{def_of_u}
  u_\pm^2 \equiv \frac{3}{\Lambda_\pm}\left(\frac{\epsilon}{3\sigma} \mp 2\pi\sigma\right)^2~,
\end{equation}
such that $2V = -H_\pm^2(1+u_\pm^2)R^2$, where $H_\pm^2 = \Lambda_\pm/3$.
The solution is valid in $\mathcal{M}_-$ and $\mathcal{M}_+$, with the corresponding quantities $(u_-,\Lambda_-)$ and $(u_+,\Lambda_+)$ respectively.
Equation~(\ref{posLTB}) becomes
\begin{equation}
  \frac{2V(a\bar r)}{H}\frac{\partial_t \bar r}{\bar r} = 1 - |u|\sqrt{-2V(a\bar r)-1}~.
\end{equation}
which, when rewritten as a differential equation for $V$, can be
solved by separation of variables. Solving for $\bar r$ yields
\begin{equation} \label{dS_exactsolution2}
  \bar r(t) = \sqrt{u^{-2} + \left(\exp\left(-Ht\right) -1\right)^2}H^{-1}~.
\end{equation}
For convenience we normalized the scale factor at the time $t=t_0$ when $\partial_t \bar r(t_0)=0$
and took $t_0=0$ without loss of generality.
This implies
\begin{equation}
  \bar r_0^{~\pm} = \left|\frac{\epsilon}{3\sigma} \mp 2\pi\sigma\right|^{-1}~.
\end{equation}
After $t=t_0$ the bubble accelerates and converges to $\bar r(t\rightarrow\infty) = \sqrt{1+u^2}\bar r_0$,
see Fig.~(\ref{dS_exact_trajectories2}).
We conclude with the remark that~(\ref{dS_exactsolution2}) reduces to the solution found in~(\ref{dS_exactsolution1}) in the limit of $G\rightarrow 0$.

\subsubsection{Numerical solution in de~Sitter/LTB spacetime}

The goal of this section is to understand the influence of ambient inhomogeneities on the motion of the bubble wall.
The first step into the numerics of the bubble is solving the dynamics of the background.

For an intuitive approach we define $2E(r) = -k(r)r^2$ where the profile $k$ may be interpreted as the local spatial curvature.
Then, the coordinate size of the spatial section, if finite, is determined by $k\left(r_\mathrm{max}\right)r_\mathrm{max}^2 = 1$.
In addition, we have to specify the initial value of the scale factor $a_0(r)\equiv a(t_0,r)$
which will in general depend on the radial coordinate.
Instead, one may equivalently choose the initial dust density $\rho_0(r)\equiv \rho(t_0,r)$, which
defines $a_0(r)$ via equation~(\ref{eqmodust}).
Thus, in coordinates where $M(r) = \frac{4\pi}{3}Ar^3$,
spatial inhomogeneity of the LTB spacetime is incorporated in the functions $k$ and $\rho_0$.
In the limit where $k$ and $\rho_0$ are constant the model becomes homogeneous.

After both functions have been specified, the partial differential equation~(\ref{eqmoLTB}) can directly be integrated at each $r$.
When a solution is found its validity has to be checked.
If not $\partial_r(ar)>0$, the weak energy condition $\rho \geq 0$ is violated by the occurrence of a shell-crossing singularity.
This actually restricts the curvature profile $k$, since, if too steep,
it disturbs the background massively such that the dust density will violate the weak energy condition at some time.
We evolve the system until the background space has expanded for about 4 efolds.

The initial time of the analysis is $t=t_0$ where $\partial_t \bar r(t_0)=0$.
$t_0$ will be referred to as the time of nucleation of the bubble.
The nucleation radius is determined by the parameters $A, \Lambda_+, \Lambda_-, \sigma_0$.
It can be inferred from equation~(\ref{posLTB}) which we rearrange to
\begin{equation} \label{nuclradius}
  \frac{1}{\bar r_0^2} = k(\bar r_0) +a_0^2(\bar r_0)\left(\frac{\epsilon_0}{3\sigma_0}-2\pi\sigma_0 \right)^2~.
\end{equation}
Note that the initial difference in energy density $\epsilon_0$ includes the initial dust density $\rho_0$,
which can be a function of $\bar r_0$, too.
This equation illustrates the route that we will follow.
There are two possibilities, via the functions $k$ and $\rho_0$, to introduce inhomogeneity in the LTB model.
These two cases are considered independently, meaning that one of the two terms on the right hand side will be independent of $r$.
In case there is more than one solution to the equation the smallest positive value will be taken.

Note also that the proper kinetic energy of the bubble at nucleation is proportional to
$  \dot R^2 \vert_{t=t_0} = H_0^2R_0^2$,
where $H_0 \equiv \frac{\partial_t a}{a}(t_0,\bar r_0)$ and $R_0=a_0(\bar r_0)\bar r_0$.

\paragraph{Homogeneous limit}

The first thing we will explore is not the effect of inhomogeneity,
but simply what happens when the bubble nucleates in a background where, in addition to vacuum energy density, some dust is present.
To keep things simple we refrain from adding any curvature at this stage and set $k=0$ and $a_0(r)=1$.
The spatial sections of LTB spacetime become homogeneous and reduce to the FLRW limit.
Henceforth, we will always assume that at nucleation dust density shall exceed exterior vacuum energy,
and for definiteness we choose $8\pi A/3=10^{-4}$ and $\Lambda_+/3=10^{-5}$ in accordance with~\cite{Fischler07}.

It is important to note that this setup already has a significant effect on the evolution of the bubble.
Whereas a bubble that nucleates in vacuum always begins to expand, this is no longer guaranteed as soon as
considerable amounts of matter are present in the environment. This can be seen by the following argument.
The force which accelerates the shell has two contributions: one from the surface tension, which is always
directed inwards, and one from the pressure difference between the interior and exterior fluid.
In the case where both fluids are mere cosmological constants it is easy to show that the pressure
force can sustain the surface tension and will push the shell outwards. However, if the exterior fluid
is mainly composed of pressureless dust, at some point surface tension will outrun pressure support
and the bubble will be forced to collapse.

\begin{SCfigure}[5][tp]
\centering
  \psfrag{r/e}[b][b]{\small $\rho / \epsilon_{\mathrm{vac}}$}
  \psfrag{6pGs2/e}[t][t]{\small $6 \pi \sigma^2 / \epsilon_{\mathrm{vac}}$}
  \psfrag{collapsing}[b][b]{\small \textbf{contracting}}
  \psfrag{expanding}[t][t]{\small \textbf{expanding}}
  \psfrag{forbidden}[b][b]{\small \textbf{forbidden}}
  \psfrag{smallest possible rL}[b][b]{\footnotesize matter dominated universes possible}
  \psfrag{vacuum dominated universes}[t][t]{\footnotesize all universes vacuum dominated}
  \includegraphics[width=0.55\textwidth]{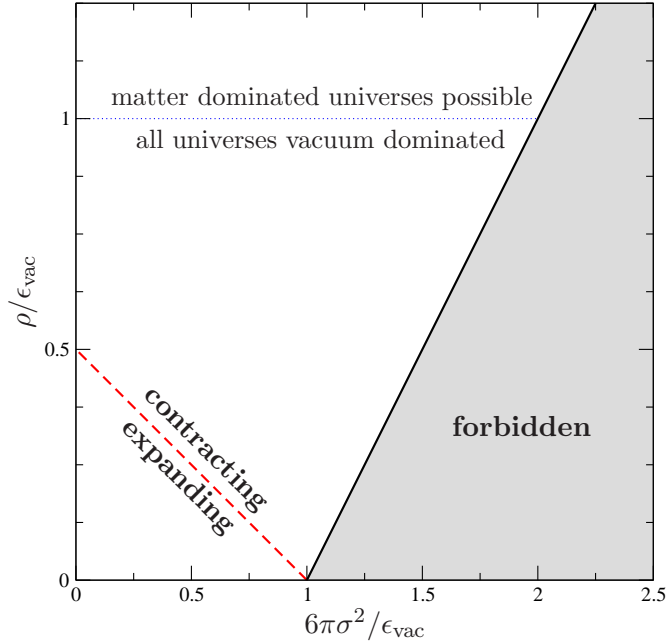}
  \caption{\label{sigmabounds}
    This plot characterizes the early-time behavior of a vacuum bubble after it was nucleated at rest in
    the comoving frame of an exterior flat FLRW spacetime with dust and cosmological constant. With
    respect to an exterior comoving observer, the bubble shows different behavior in different regions
    of the $\sigma^2$-$\rho$-plane, which is drawn in units of $\epsilon_{\mathrm{vac}} \equiv \left(\Lambda_+ - \Lambda_-\right) / 8 \pi$.
    The shaded region is forbidden for our choice of junction because $\sigma$ there violates the bound (\ref{bound_sigma}).
    If the energy density of dust $\rho$ is chosen above the dashed red line, the bubble starts to contract.
    This includes all matter dominated universes $\rho > \rho_{\mathrm{vac}} \equiv \Lambda_+ / 8 \pi$
    since we assume $\Lambda_+ > \Lambda_- \geq 0$, which implies $\rho_{\mathrm{vac}} \geq \epsilon_{\mathrm{vac}}$.
    Below the dashed red line the bubble starts to expand into the ambient spacetime. This includes all
    vacuum de Sitter spacetimes since they are found on the line $\rho = 0$. Below the dotted blue line, all
    universes are vacuum dominated.}
\end{SCfigure}

To make this statement more quantitative we can take another derivative of eq.~(\ref{posLTB}) and examine the behavior
of the bubble when $\partial_t \bar{r} = 0$. In the homogeneous limit one finds
\begin{equation}
\left. \partial_t^2 \bar{r}\right|_{\partial_t \bar{r} = 0}~=~\frac{1}{a} \left(\frac{\Lambda_+ - \Lambda_-}{24 \pi \sigma} - 2 \pi \sigma - \frac{2 \rho}{3 \sigma}\right)~.
\end{equation}
The sign of $\partial_t^2 \bar{r}$ depends on how $\sigma^2$ and $\rho$ compare to
the latent heat of the vacuum, $\epsilon_{\mathrm{vac}} \equiv \left(\Lambda_+ - \Lambda_-\right) / 8 \pi$.
The bubble can only expand into the ambient spacetime if $\rho$ is not too large. In particular,
one can never have an expanding bubble during a matter dominated phase\footnote{Note that this was not
at all an issue in section 2, since background spacetime was assumed spatially homogeneous and any 
dust would therefore permeate the bubble. In the present setup, however, the interior is assumed
to be completely empty except for
a possible cosmological constant. In this sense, the issue is one of initial conditions.}.
This strongly limits the possibility
to study the propagation of bubbles into inhomogeneous matter with the current approach, since
the exterior spacetime has to be vacuum dominated in order to allow the bubble to propagate towards the
inhomogeneity in the first place. We have summarized the behavior of freshly nucleated bubbles of vacuum
within a dust environment in Fig.~\ref{sigmabounds}.

\begin{figure}
\begin{center}
\begin{tabular}{lr}
  \psfrag{y}[][]{\scriptsize{$H_0\bar r$}}
  \psfrag{x}[][]{\scriptsize{$\sqrt{\Lambda_+/3}~t$}}
  \includegraphics[width=0.47\textwidth]{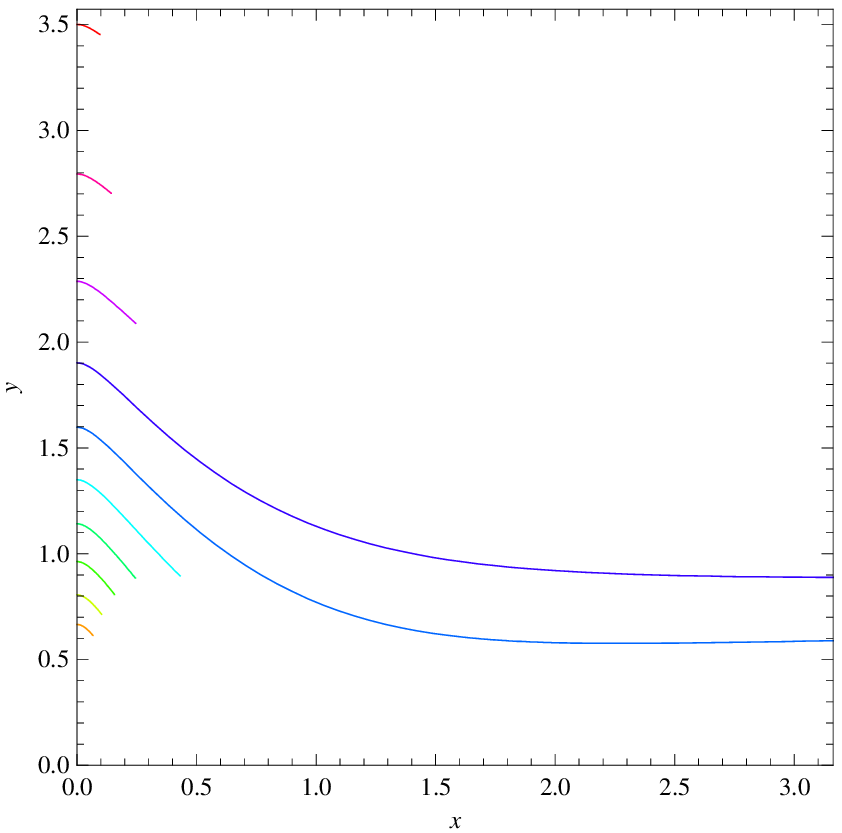}
&
  \psfrag{y}[][]{\scriptsize{$\Lambda_-/\Lambda_+$}}
  \psfrag{x}[][]{\scriptsize{$4\pi\sigma_0/H_0$}}
  \includegraphics[width=0.47\textwidth]{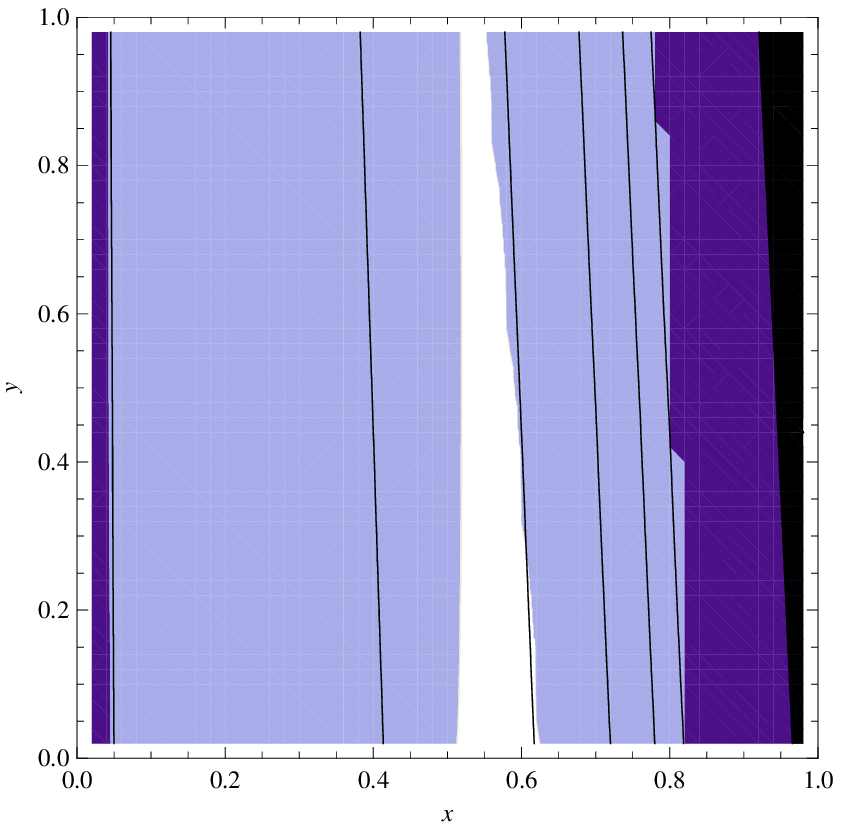}
\end{tabular}
  \caption{\label{figures_LTB_flat}
    Results obtained for the homogeneous limit of the LTB model with $k=0$ and $a_0(r)=1$.
    Left figure:
    Trajectories of the bubble wall for several values of the surface tension~
    $0.3H_0\le 4\pi\sigma_0 \le 0.75H_0$ and~$\Lambda_-=0.1\Lambda_+$.
    The fate of the bubble depends on the nucleation size.
    Small bubbles contract in LTB coordinates until their proper kinetic energy becomes zero, i.e.~$\dot R = 0$
    and the equation of motion becomes imaginary.
    The greater the bubble the more likely it sustains kinetic energy until the background is dominated by~$\Lambda_+$
    and it converges to some finite coordinate radius.
    Right figure:
    Fate of a bubble in dependence of the parameters~$\sigma_0,\Lambda_-$.
    We looked in the region of parameter space where the nucleation radius of the bubble is within $0.1<H_0\bar r_0<5$.
    The black lines are lines of constant kinetic energy and nucleation radius~$H_0\bar r_0=(0.1,1,2,3,4,5)$.
    The black shaded region is considered to be unphysical since equation~(\ref{bound_sigma}) is not fullfilled there already initially.
    In the region shaded light blue, bubbles either contract until~$\dot R = 0$, or they hit the geometrical bound~(\ref{bound_sigma}).
    Neither occurs in the white region in which bubbles \lq survive\rq~and come to rest at a finite coordinate radius.}
\end{center}
\end{figure}

We numerically calculated the trajectories of bubbles in a dust dominated background. As expected
by the argument above, in contrast to the de~Sitter case, bubbles contract as seen from the exterior perspective
(Fig.~\ref{figures_LTB_flat}, left).
In fact some bubbles contract so fast that the growth of their physical size is decelerated, i.e. their proper kinetic energy decreases.
Our results show that it will even decrease to zero for small bubbles with $H_0\bar r_0 \lesssim 1$.
In this case the proper kinetic energy becomes imaginary and the evolution of the bubble had to be stopped.
When $\bar r_0\gtrsim H_0^{-1}$ they retain some proper kinetic energy but nevertheless shrink in exterior coordinates and converge to a coordinate radius which is smaller than the coordinate radius of nucleation.
For bubbles larger than $2H_0^{-1}$, even if they fulfill the bound~(\ref{bound_sigma}) initially, the
dust density in the background drops faster than the surface tension of the bubble such that the bound will be violated soon.

Note also that after $\sqrt{\Lambda_+/3}~t_\mathrm{eq}\simeq 0.4$ the bubble moves on a vacuum dominated background.
The contour plot in \ref{figures_LTB_flat} illustrates the fate of a bubble in dependence of the parameters $(\Lambda_0,\sigma_0)$.

\paragraph{Inhomogeneous dust density}

\begin{figure}
\begin{center}
\begin{tabular}{lr}
  \psfrag{y}[][]{\scriptsize{$\sqrt{\Lambda_+/3}~\bar r$}}
  \psfrag{x}[][]{\scriptsize{$\sqrt{\Lambda_+/3}~t$}}
  \includegraphics[width=0.47\textwidth]{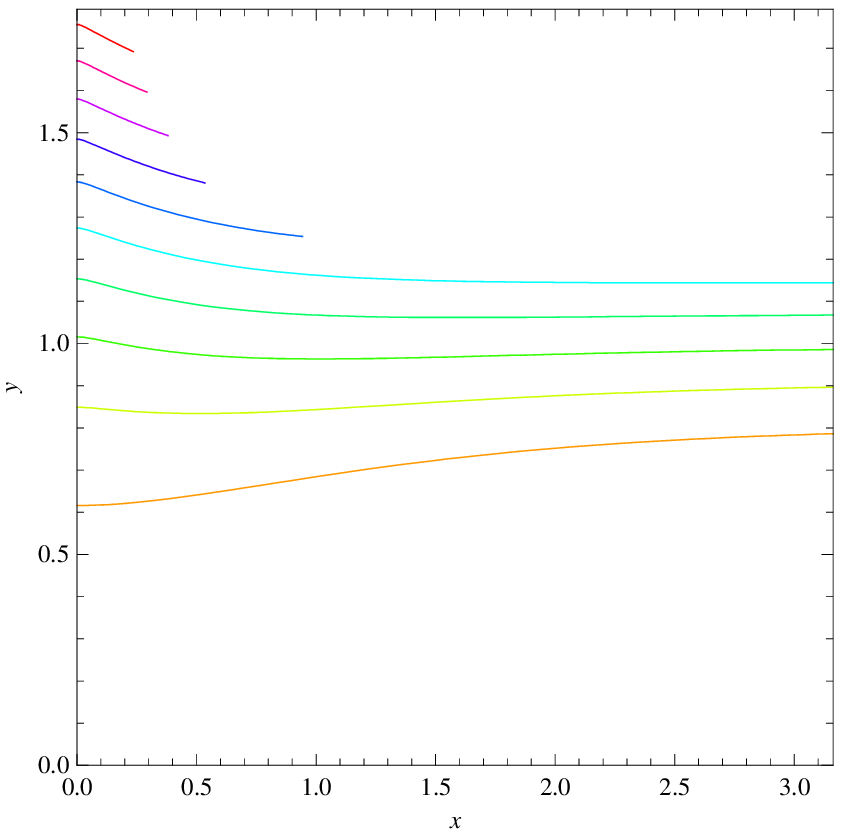}
&
  \psfrag{y}[][]{\scriptsize{$\rho(t,\bar r)/A$}}
  \psfrag{x}[][]{\scriptsize{$\sqrt{\Lambda_+/3}~t$}}
  \includegraphics[width=0.47\textwidth]{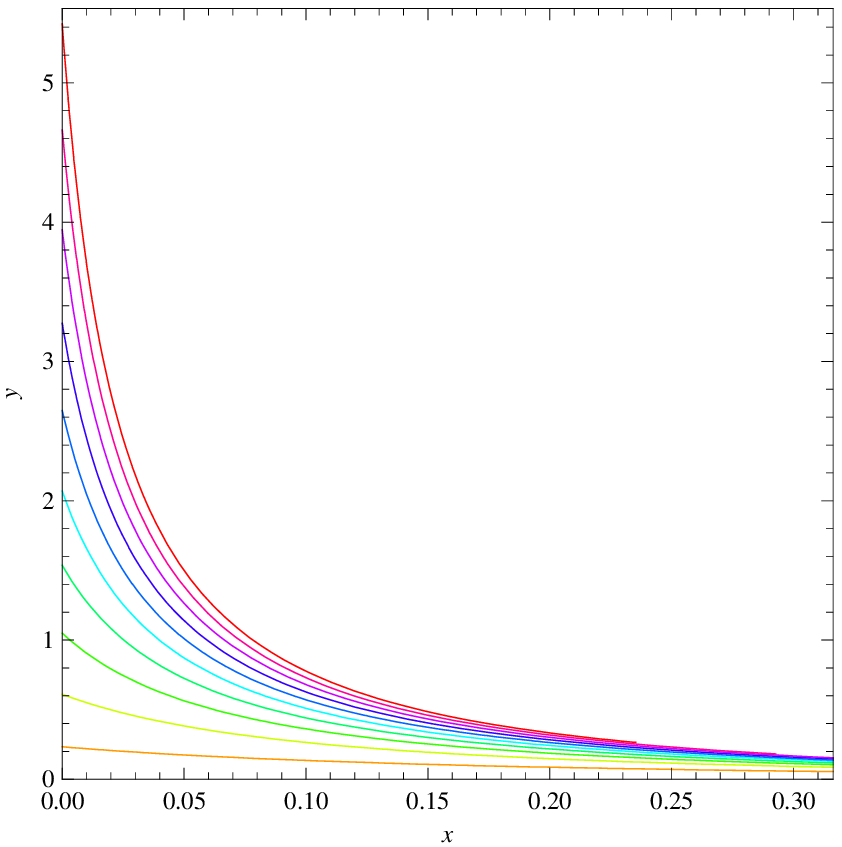}\\
\end{tabular}
  \caption{\label{figures_LTB}
    Results obtained for the inhomogeneous LTB model with the dust profile~(\ref{dustprofile}).
    The left figure shows the trajectory of bubbles with initial surface tension
    $0.7\sqrt{\Lambda_+/3}\le 4\pi\sigma_0 \le 1.6\sqrt{\Lambda_+/3}$ and~$\Lambda_-=0.1\Lambda_+$.
    Smaller bubbles nucleate in a region where vacuum energy dominates over dust density and therefore expand.
    Bubbles with $\bar r_0 \simeq r_A$ contract because they are already in a dust dominated background.
    The right figure shows the dust density at the position of the bubble. Although the dust profile~(\ref{dustprofile})
    radially increases, the expansion of the background dilutes matter efficiently such that expanding bubbles
    effectively propagate in a decreasing profile.
  }
\end{center}
\end{figure}

We now explore possible effects of inhomogeneity by the 
introduction of an initial dust distribution $\rho_0(r)$. The bubble may nucleate either in an overdense or in an underdense region of space.
It is not very revealing to consider a radially decreasing dust profile,
because the dust will be rarefied by the expansion of the background anyway.
Therefore we will have a look at radially increasing dust profiles only.
For a given $\rho_0(r)\,$, the initial scale factor is determined by
\begin{equation}
  a_0^3(r) = \frac{3A}{r^3}\int \frac{r^2}{\rho_0(r)}\rmd r.
\end{equation}
We will consider the profile
\begin{equation}\label{dustprofile}
  \rho_0(r) = A r^3/r_A^3 
\end{equation}
with $r_A=\left(\Lambda_+/3\right)^{-1/2}$.
We expect that small bubbles with $\bar r_0 \ll r_A$ will begin to expand initially because they will find themselves in a vacuum dominated background, whereas large bubbles with $\bar r_0 \gtrsim r_A$ are in a matter dominated background with their subsequent evolution being
much like in the homogeneous limit discussed before.
Although the initial dust profile is increasing, expanding bubbles propagate into regions of lower density due to the expansion of the background, see Fig.~\ref{figures_LTB}.

\paragraph{Inhomogeneous curvature}

The other possibility is to incorporate inhomogeneity in the neighborhood of the bubble via a curvature profile $k$.
However, in view of the results obtained in the homogeneous limit, we state that the bubble will hardly be able propagate into that inhomogeneity because it will shrink as long as the background is matter dominated.
Even if the condition that the bubble nucleates comovingly is relaxed, such that the bubble may have $\partial_t \bar r(t_0)>0$, deceleration is large enough to make the bubble contract almost immediately.
Therefore it seems that studying the effect of curvature inhomogeneity within this approach is hardly feasible.
Note that this result appears to be in contrast to what has been obtained in \cite{Fischler07}.

\begin{figure}
\begin{center}
\begin{tabular}{lr}
  \psfrag{Hr}[][]{\scriptsize{$\sqrt{\Lambda_+/3} r$}}
  \psfrag{y}[][]{\scriptsize{$R_\mathrm{cr}\sqrt{k}$}}
  \includegraphics[width=0.47\textwidth]{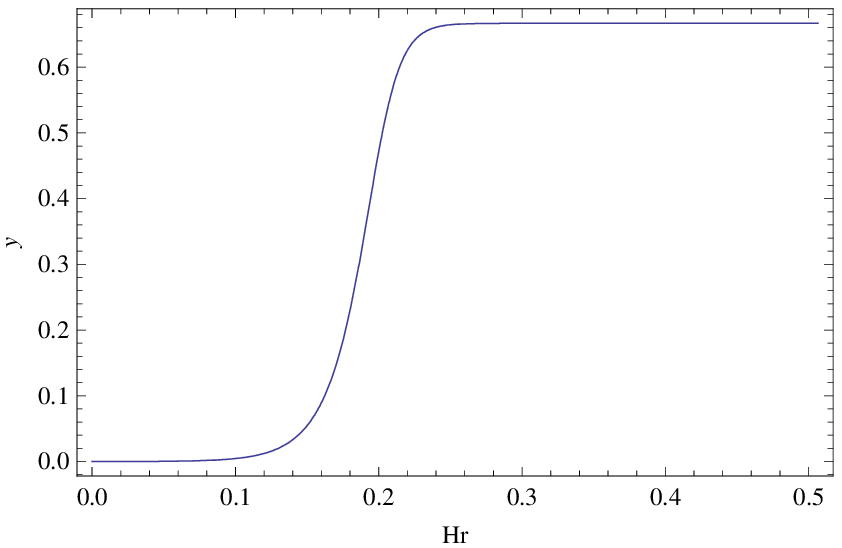}
&
  \psfrag{y}[][]{\scriptsize{$\sqrt{\Lambda_-/3}~\bar r$}}
  \psfrag{x}[][]{\scriptsize{$\sqrt{\Lambda_-/3}~t$}}
  \includegraphics[width=0.47\textwidth]{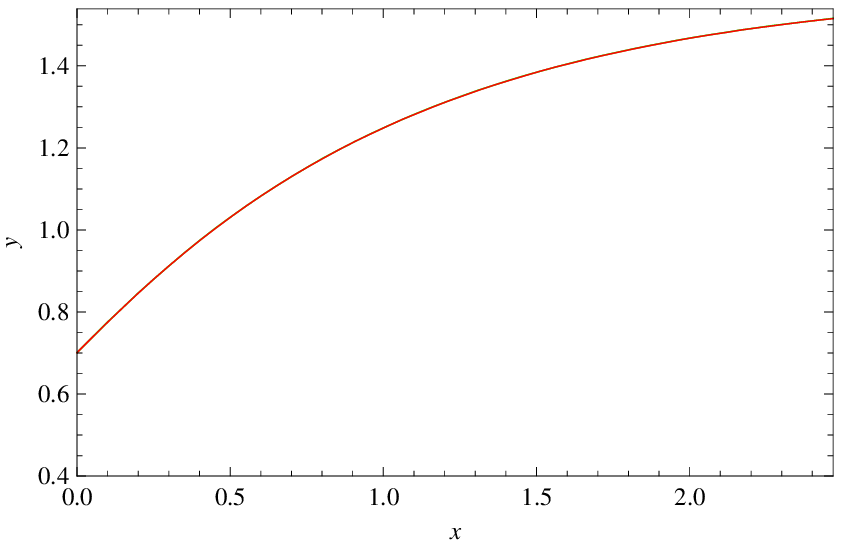}\\

  \psfrag{y}[][]{\scriptsize{$\sqrt{\Lambda_+/3}~\bar r$}}
  \psfrag{x}[][]{\scriptsize{$\sqrt{\Lambda_+/3}~t$}}
  \includegraphics[width=0.47\textwidth]{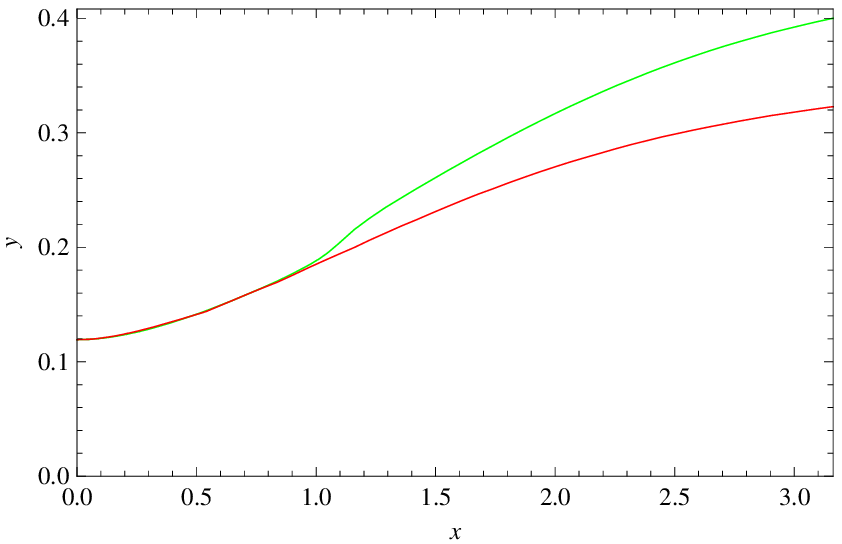}
&
  \psfrag{y}[][]{\scriptsize{$\sigma/\sigma_0$}}
  \psfrag{x}[][]{\scriptsize{$\sqrt{\Lambda_-/3}~t$}}
  \includegraphics[width=0.47\textwidth]{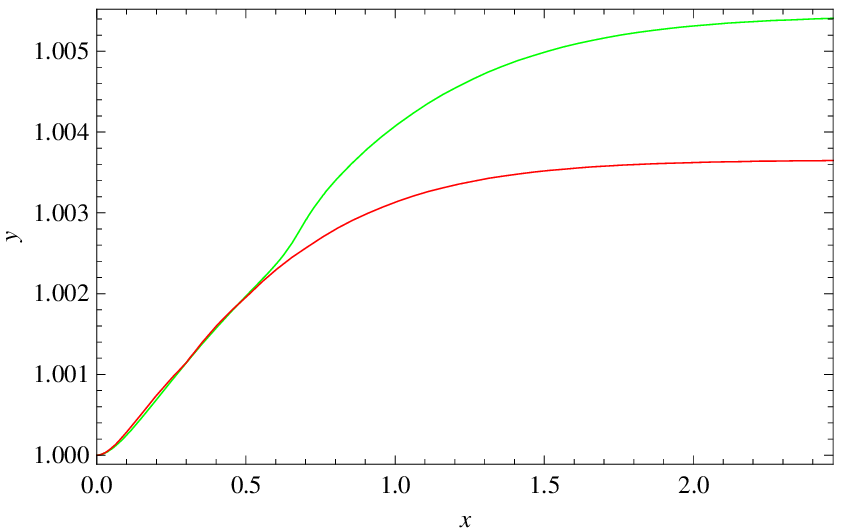}
\end{tabular}
  \caption{\label{figures_LTB_k}
    Results obtained for a bubble propagating in an inhomogeneous background (curvature inhomogeneity \textit{and} inhomogeneity in the initial dust profile). Upper left:
    the curvature profile~(\ref{eq_curvature_profile}) with $\alpha_1 = 1.5, \alpha_2 = 0.02$ and $\alpha_3 = 0.2$.
    Lower left:
    Bubble trajectories in the comoving LTB coordinates in a flat background (red) compared to the curved background (green).
    The trajectories are affected significantly in these coordinates.
    However, if one converts the trajectories to the interior de~Sitter coordinates the effect practically vanishes (upper right).
    Nevertheless, there remains an effect on the surface tension of the bubble (lower right).}
\end{center}
\end{figure}

Nevertheless, to see what happens when a bubble enters a curvature inhomogeneity we make use of a result obtained previously.
In the last section it was shown that bubbles which nucleated in a vacuum dominated region expanded initially.
Now, we add some curvature in \lq front\rq~of the bubble and have a look what happens if the bubble encounters that inhomogeneity and
whether there is a difference compared to the corresponding solution in the flat background.
The curvature profile is given by
\begin{equation} \label{eq_curvature_profile}
  k(r) = \frac{1}{2\left(\alpha_1 \mathcal{R}_\mathrm{cr}\right)^2}
         \left(1+\tanh\left( \frac{\sqrt{\Lambda_+/3}~r-\alpha_3}{\alpha_2} \right)\right)~,
\end{equation}
where $\mathcal{R}_\mathrm{cr} \equiv (4 \pi A \sqrt{\Lambda_+})^{-1/3}$. Taking the dust profile from the last section, we fix the initial surface tension and vacuum energy of the bubble to $\sigma_0=0.6\sqrt{\Lambda_+/3}, \ \Lambda_-=0.1\Lambda_+$.
The evolution of the bubble is significantly affected in the exterior perspective.
However, this may be just a coordinate effect and when looked at the trajectories in the interior coordinates the effect practically vanishes. Nevertheless, there remains an effect in the surface tension of the bubble, see Fig.~\ref{figures_LTB_k}.

\subsubsection{Numerical solution in de~Sitter/FLRW spacetime}

In this section, the evolution of the bubble on a de~Sitter/FLRW background will be considered.
The FLRW part is supposed to contain vacuum energy and a perfect fluid which undergoes a phase transition i) from~$w=1/3$ to~$w=-1$,
or ii) vice versa.
The FLRW dynamics~(\ref{eqmoFLRW}) and~(\ref{conFLRW}) will be solved numerically with
the initial values~$a(t_0)=1$,~$8\pi\rho_0/3=10^{-4}$ and $\Lambda_+/3=10^{-5}$.
Again~$t_0=0$ shall represent the exterior time coordinate at which~$\partial_t \bar r(t_0)=0$.
The phase transition is supposed to occur at~$t_\mathrm{pt}=0.5H_0^{-1}$, and we set the width $\gamma_\mathrm{pt}^{-1} = 1 \ll H_0^{-1}$ in order to model a nearly instantaneous transition.
After these dynamics have been established we consider the evolution of the bubble.

\begin{figure}
\begin{center}
\begin{tabular}{lr}
  \psfrag{y}[][]{\scriptsize{$H_0\bar r$}}
  \psfrag{x}[][]{\scriptsize{$H_0 t$}}
  \psfrag{W}[][]{\scriptsize{\fbox{$w=1/3 \rightarrow w=-1$}}}
  \includegraphics[width=0.47\textwidth]{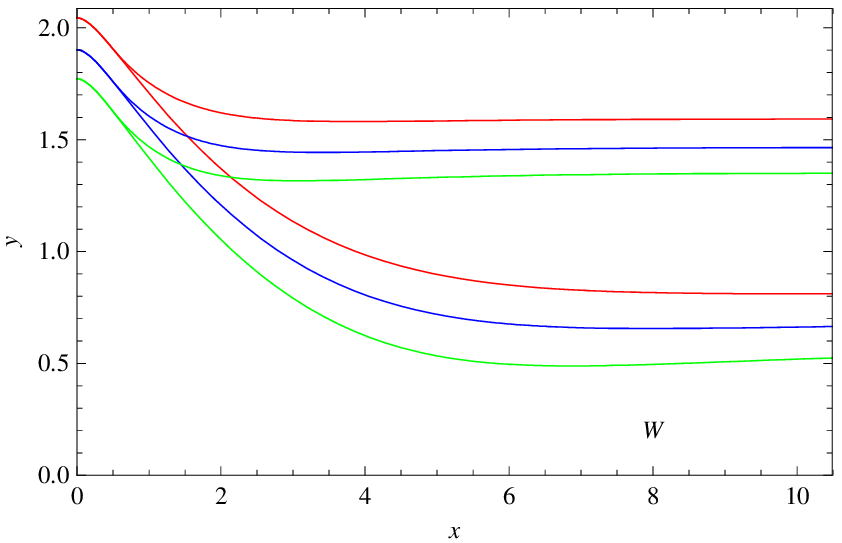}
&
  \psfrag{y}[][]{\scriptsize{$H_0\bar r$}}
  \psfrag{x}[][]{\scriptsize{$H_0 t$}}
  \psfrag{W}[][]{\scriptsize{\fbox{$w=-1 \rightarrow w=1/3$}}}
  \includegraphics[width=0.47\textwidth]{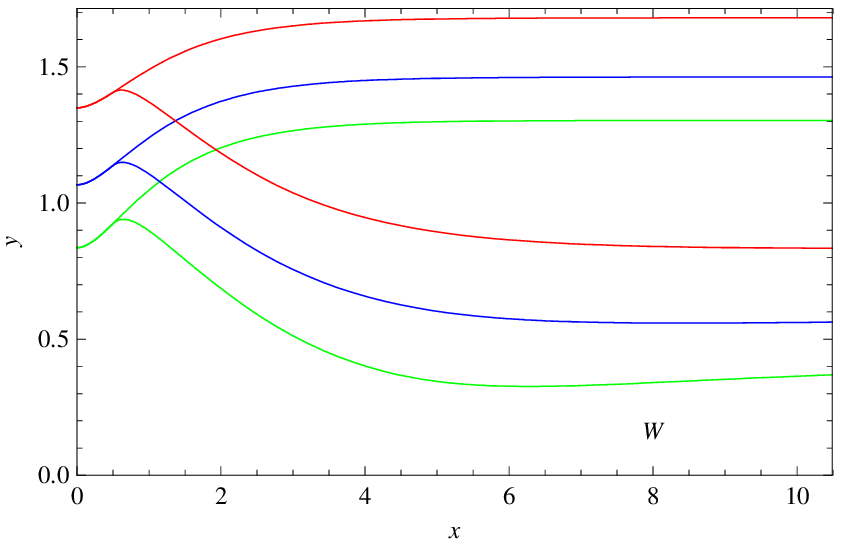}\\

  \psfrag{y}[][]{\scriptsize{$\sqrt{\Lambda_-/3}~\bar r$}}
  \psfrag{x}[][]{\scriptsize{$\sqrt{\Lambda_-/3}~t$}}
  \psfrag{W}[][]{\scriptsize{\fbox{$w=1/3 \rightarrow w=-1$}}}
  \includegraphics[width=0.47\textwidth]{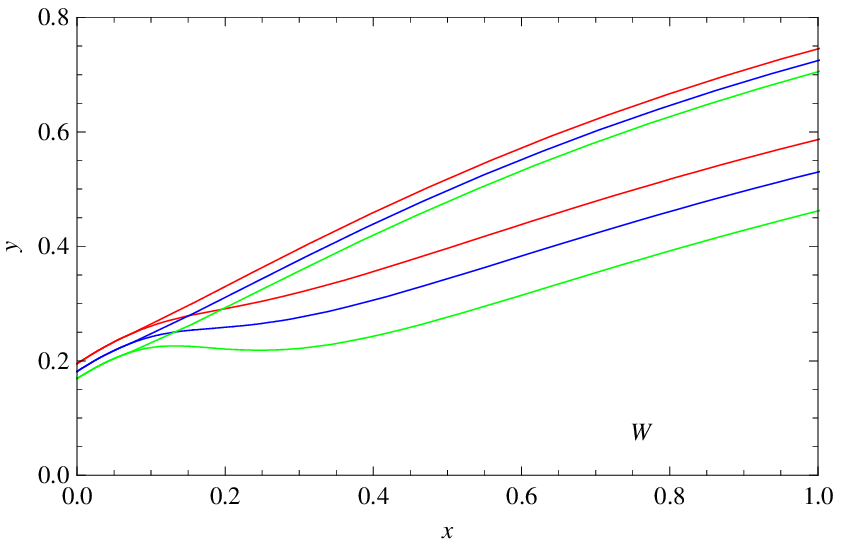}
&
  \psfrag{y}[][]{\scriptsize{$\sqrt{\Lambda_-/3}~\bar r$}}
  \psfrag{x}[][]{\scriptsize{$\sqrt{\Lambda_-/3}~t$}}
  \psfrag{W}[][]{\scriptsize{\fbox{$w=-1 \rightarrow w=1/3$}}}
  \includegraphics[width=0.47\textwidth]{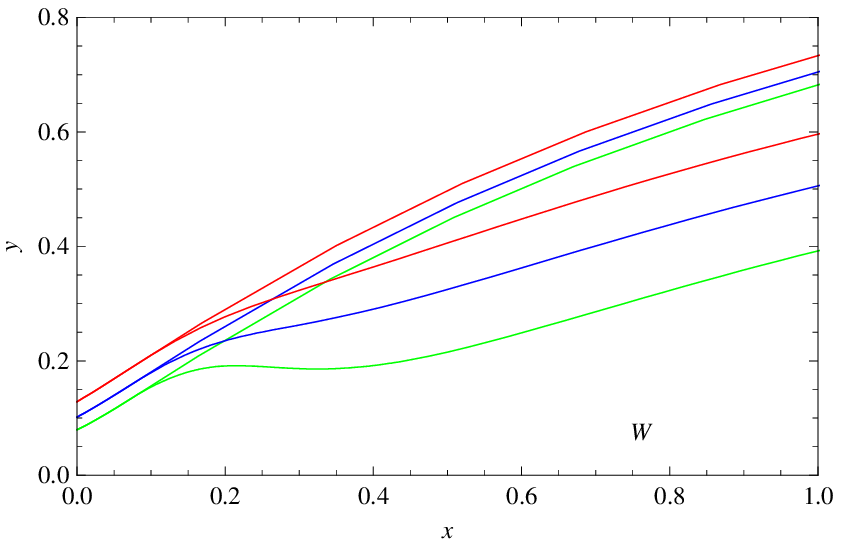}\\

  \psfrag{y}[][]{\scriptsize{$\sigma/\sigma_0$}}
  \psfrag{x}[][]{\scriptsize{$\sqrt{\Lambda_-/3}~t$}}
  \psfrag{W}[][]{\scriptsize{\fbox{$w=1/3 \rightarrow w=-1$}}}
  \includegraphics[width=0.47\textwidth]{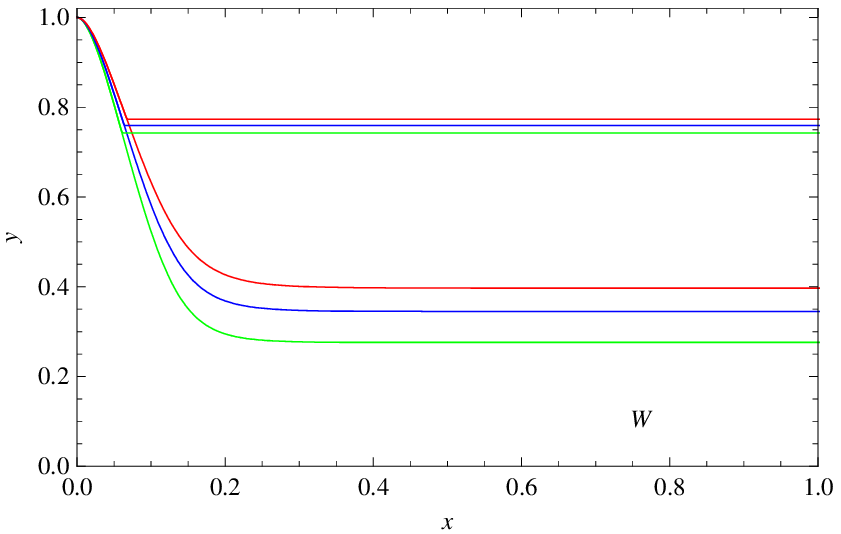}
&
  \psfrag{y}[][]{\scriptsize{$\sigma/\sigma_0$}}
  \psfrag{x}[][]{\scriptsize{$\sqrt{\Lambda_-/3}~t$}}
  \psfrag{W}[][]{\scriptsize{\fbox{$w=-1 \rightarrow w=1/3$}}}
  \includegraphics[width=0.47\textwidth]{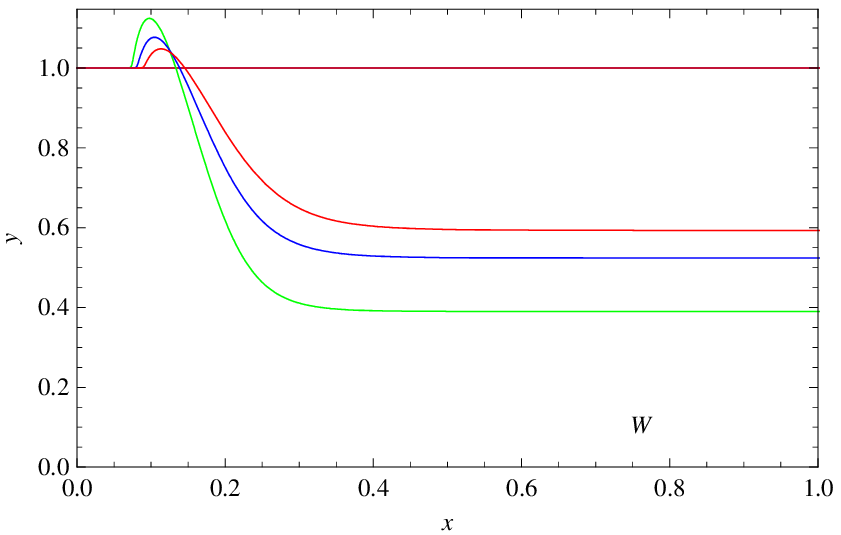}
\end{tabular}
  \caption{\label{figures_FLRW}
    Evolution of bubbles in the de~Sitter/FLRW spacetime for surface tensions $0.4H_0 \leq 4\pi\sigma \leq 0.6H_0$
    and $\Lambda_- = 0.1\Lambda_+$.
    The left column shows the results of a transition from $w=1/3$ to $w=-1$ and the right column from $w=-1$ to $w=1/3$
    compared, respectively, to their counterparts where the equation of state remains constant.
    The upper plots show the trajectory of the bubble in exterior coordinates, while the mid plots show the trajectories
    as seen from an interior observer.
    Unlike for the inhomogeneous background, we see that the exterior phase transition indeed leaves a sizeable effect
    on the trajectory of the bubble. In addition, the most prominent effect is again on the surface tension.}
\end{center}
\end{figure}

The nucleation radius of the bubble can still be obtained from equation~(\ref{nuclradius}) with $k=0$.
Of course, an immediate effect of the phase transition on the trajectory of the bubble can be seen in the exterior coordinates.
In case i) the contraction of the bubble is stopped when the background becomes vacuum dominated,
whereas in case ii) the expansion of the bubble reverses to contraction due to the phase transition.
However, contrary to the inhomogeneous background discussed before, the effect is still present when the trajectory is expressed in the coordinates of an interior observer, see Fig.~\ref{figures_FLRW}.
However, the most prominent effect still can be seen in the surface tension of the bubble.

\section{Conclusions}

First-order phase transitions, as may have occurred multiple times
in the early universe, proceed by spontaneous nucleation of bubbles.
In order to answer the question on how much information about the
initial state \lq survives\rq~the phase transition, it is important to
understand bubble nucleation and propagation beyond the assumption of a trivial initial state. This
question may indeed be relevant for certain cosmological scenarios,
such as the chain inflation proposal. In this particular scenario, a
series of first-order phase transitions proceeds very quickly, such
that the time in between two transitions is too short for the universe
to dilute all inhomogeneities and thermal radiation produced in
each transition.

We have seen that the dynamics of the background spacetime affects
the calculation of semiclassical tunneling probabilities. A simple
comparison of time scales helps to decide if this effect is relevant.
In cases where the bubble crossing time (the nucleation radius of
the bubble divided by the speed of light) is much smaller than any time
scale of the background geometry, we found that it is well justified
to use tunneling rates obtained from field theory on Minkowski spacetime.
Noting that this tunneling probability drops exponentially with
increasing nucleation radius, it seems reasonable to assume that the
bubbles in fact nucleate at tiny radii whenever we demand an appreciable
tunneling rate which can lead to an onset of the phase transition
before the universe reaches (approximately) a vacuum state. Hence, using
Minkowski spacetime as an approximation appears self-consistent in many
cases. However, if the expansion rate of the universe is very large,
there may be cases where the effect of background evolution on tunneling
rates can be important. In the context of the chain inflation proposal,
we expect this to be the case very early on during a radiation dominated
phase short after a previous phase transition.

The existence of a particle horizon in a radiation dominated FLRW
universe renders the nucleation of bubbles larger than this horizon
impossible. This means that one has to have better knowledge of the
spacetime near the singularity, including details about reheating and
any cosmology preceding the radiation era, in
order to avoid this \lq horizon problem\rq . In other words, the pure
radiation dominated FLRW universe is no useful approximation for a
semiclassical calculation of nucleation rates of bubbles larger than
the particle horizon, since those rates would be sensitive to the
cosmology at the Big Bang.

Concerning the subsequent evolution of comovingly nucleated vacuum bubbles in non-vacuum backgrounds,
we have seen that already the presence of homogeneously distributed matter has a
significant influence on the bubble. Unlike in de~Sitter spacetime where the
proper kinetic energy of a bubble increases exponentially, it may decrease in the presence of
matter in the background. For small bubbles, the proper kinetic energy became zero 
and real classical solutions could not be obtained beyond this point. As has been pointed out in
\cite{Fischler07}, such bubbles do not correspond to classical configurations, but should
be interpreted as mere fluctuations.
Bubbles with greater radius have more proper kinetic energy and are able to survive until the matter
density in the background has been diluted sufficiently. We note, however, that the setup in
which we study bubble trajectories, in particular the choice of initial conditions, is somewhat ad hoc.
To settle this issue, one would have to solve the tunneling problem with matter \textit{and} gravity,
an enterprise on which we did not embark in this work.

We also studied the effect of inhomogeneities in the background on the propagation of vacuum bubbles.
The results show that the trajectory of a bubble is affected, from the point of view of an exterior observer,
when it propagates into an inhomogeneous background. Since it is not clear whether this is just a coordinate effect
we have looked at the trajectory as seen by an interior observer. This observer will hardly
see an influence of ambient inhomogeneities in the trajectory of the bubble.
Nevertheless, when looked at from the inside, there remains an effect on the surface tension of the bubble.

Furthermore, bubbles moving in an FLRW spacetime which itself undergoes a phase transition have been considered.
Similar to the inhomogeneous case, a large effect is seen in the exterior coordinates.
However, when converted to interior coordinates, there remains an appreciable effect in the trajectory of the bubble.
This raises the question whether those perturbations of the bubble wall will lead to potentially observable effects.
In the context of bubble collisions \cite{Aguirre08,Chang08}, it has been pointed out that a disturbance in the
trajectory of the bubble wall may lead to a redshift of the reheating surface and could therefore potentially be observable in the CMB.

However, the most prominent of the effects of inhomogeneity or phase transitions in the background
is found in the surface tension of the bubble. This is a consequence of the
spacetime junction approach. As soon as a bubble propagates in a matter environment the evolution of
its surface tension is determined by the demand of the background.
This means that an expanding bubble \lq collects\rq\ matter from the background while a contracting bubble
will lose the amount of energy required by the space it uncovers.
Therefore it is necessary to 
find a proper treatment of energy transfer through the bubble surface. Rather than fixing interior and
exterior spacetime ab initio, one should dynamically construct those spacetimes from initial conditions.
To solve this problem, an additional equation is needed, which arises from a proper dynamical description
of the surface tension. It should capture the physics of matter transfer across the junction hypersurface
and should probably be motivated from a field theoretical point of view. We hope to make progress in this
direction in our future work.

\ack
We thank Ben Freivogel, I-Sheng Yang, Anthony Aguirre, 
and Douglas Spolyar for helpful discussions and suggestions.
This work was supported by the German Research Foundation 
(DFG) through the Research Training Group 1147 
\lq Theoretical Astrophysics and Particle Physics\rq.

\begin{appendix}

\section{Evolution equations in de~Sitter/FLRW spacetime}

For the proposed FLRW spacetime the relevant equations can be obtained in the same way as for the LTB part
by setting $\partial_r a=0$ and $E=0$.
\begin{equation}
	ar = R, \quad {\dot t}^2 = 1 + a^2{\dot r}^2~.
\end{equation}
The potential $V$ reduces to
\begin{equation}
	2V = -\left[\frac{\Lambda_-}{3} 
		+\left(\frac{\rho}{3\sigma} +\frac{\Lambda_+ -\Lambda_-}{24\pi\sigma} +2\pi\sigma \right)^2\right]R^2~,
\end{equation}
and the equations of motion are given by
\begin{eqnarray}
  \partial_t \bar r =
    \frac{\bar r\partial_t a -\sqrt{\left(1+2V\right)\left(\left(\bar r\partial_t a\right)^2+2V\right)}}{2aV}~, \\
  \partial_t \sigma =
    \left(\rho+p\right)\frac{a\partial_t \bar r}{\sqrt{1-\left(a\partial_t \bar r\right)^2}}~.
\end{eqnarray}
These equations together with the background dynamics given by (\ref{eqmoFLRW}) and (\ref{conFLRW}) determine the bubble motion in
the de~Sitter/FLRW background.

\end{appendix}

\end{document}